\documentclass[prd,preprint,nofootinbib]{revtex4}
\usepackage{framed}
\usepackage{bbold}
\usepackage{slashed}
\usepackage{graphicx,hyperref}
\usepackage{amsmath}
\usepackage{amssymb}
\usepackage{epsfig}
\usepackage{hhline}
\usepackage{soul}
\usepackage{tabularx,pifont,multirow}
\usepackage{enumitem}
\usepackage{colordvi}
\usepackage{soul}
\usepackage{xcolor}
\usepackage{yfonts}

\newcommand{\be}{\begin{equation}}
\newcommand{\ee}{\end{equation}}
\newcommand{\bea}{\begin{eqnarray}}
\newcommand{\eea}{\end{eqnarray}}

\newcommand{\ntext}[1]{{\textcolor{black}{#1}}}

\newcommand{\CC}{\Lambda}

\newcommand{\MPl}{M_{\rm Pl}}

\definecolor{darkgreen}{rgb}{0,0.3,0.05}
\makeatletter                                                         %
\newcommand*\rel@kern[1]{\kern#1\dimexpr\macc@kerna}                  %
\newcommand*\widebar[1]{                                              %
  \begingroup                                                         %
  \def\mathaccent##1##2{                                              %
    \rel@kern{0.8}                                                    %
    \overline{\rel@kern{-0.8}\macc@nucleus\rel@kern{0.2}}             %
    \rel@kern{-0.2}                                                   %
  }                                                                   %
  \macc@depth\@ne                                                     %
  \let\math@bgroup\@empty \let\math@egroup\macc@set@skewchar          %
  \mathsurround\z@ \frozen@everymath{\mathgroup\macc@group\relax}     %
  \macc@set@skewchar\relax                                            %
  \let\mathaccentV\macc@nested@a                                      %
  \macc@nested@a\relax111{#1}                                         %
  \endgroup                                                           %
}                                                                     %
\makeatother                                                          %

\begin{document}

\preprint{\leftline{KCL-PH-TH/2021-{\bf 16}}}

\title{Inflationary physics and transplanckian conjecture in the\\ Stringy Running-Vacuum-Model: \\  from the
phantom vacuum to  the true vacuum  \vspace{2cm}}

\author{{\bf Nick~E.~Mavromatos$^{a}$} and \vspace{0.5cm} {\bf Joan~Sol\`a~Peracaula$^b$}}

\affiliation{$^a$Theoretical Particle Physics and Cosmology Group, Physics Department, King's College London, Strand, London WC2R 2LS.\\
\vspace{0.5cm}\\
 $^{b}$Departament de F\'\i sica Qu\`antica i Astrof\'\i sica, \\ and \\ Institute of Cosmos Sciences (ICCUB), Universitat de Barcelona, Av. Diagonal 647 E-08028 Barcelona, Catalonia, Spain. \\
 \vspace{0.5cm}
}


\begin{abstract}
\vspace{0.05cm}

In previous works we have embedded the Running Vacuum Model (RVM) of Cosmology in the framework of string theory.
Specifically, we considered a string-inspired  Cosmology with  primordial
gravitational waves (GW) and gravitational anomalies, which
were argued to lead, via appropriate condensation during the very-early-universe era, to dynamical inflation, of RVM type, without the need for extra inflaton fields.
A crucial role for the associated slow-roll nature of the inflationary era was played by the fundamental axion field that exists in the gravitational multiplet of strings,  viz. the Kalb-Ramond (KR) axion.
In this paper, we study further this model and demonstrate several novel facts, completing our previous studies. We clarify the different roles played by the background KR axions,
which constitute a form of stiff matter that dominates a pre-RVM-inflationary epoch of the Universe.  We show that the KR axion when combined with
the gravitational Chern-Simons contribution obey together a peculiar equation of state  $p=-\rho$,
with negative energy density $\rho < 0$, which we call `phantom vacuum'.
Eventually, this state transmutes into the standard vacuum state thanks to  the contribution from the gravitational Chern-Simons  condensate, which makes the total $\rho_{\rm total} =-p_{\rm total} >0$.
At this point the RVM vacuum picture emerges naturally, with an overall equation of state of true vacuum  type.  We  find  that our scenario  is consistent with the trans-Planckian censorship hypothesis,  given that the
rate of change of the KR axion  takes only sub-Planckian values.
We also discuss some explicit scenarios on the creation of  GWs within the context of  supergravity models that could be embedded in our string theory framework.
{Finally, we argue  how  the  RVM in our context can help to alleviate the current tensions of the $\CC$CDM.}

\end{abstract}

\maketitle




\tableofcontents

\section{Introduction \label{sec:intro}}

In a series of recent works~\cite{bms1,bms2,bms3,msb} ({see \cite{ms21} for a review}),  we have presented an embedding of the running vacuum model (RVM) of cosmology~\cite{ShapSol1,Fossil2008,ShapSol2,JSPRev2013,JSPRev2015} into string theory by considering a low-energy effective gravitational theory
for the early universe with gravitational anomalies and primordial gravitational wave perturbations in a very early epoch.
A crucial assumption in our approach was that only degrees of freedom (d.o.f.) from the
{\it massless} bosonic  gravitational multiplet of the (super)string were
present in the respective effective action during that era.  {These
d.o.f.  are  the dilaton, gravitons and antisymmetric tensor (Kalb-Ramond
(KR)) fields, the latter being dual in four dimensions to a {\it massless} pseudoscalar d.o.f.,  $b$ (KR axion)}.  This allowed for a consistent
effective theory in the presence of gravitational Chern-Simons (gCS)
type terms, whose CP-violating coupling to the KR axions implies an exchange of energy between those fields and gravity. {By assuming constant dilatons, which are consistent string backgrounds, it was shown that condensation of primordial gravitational waves (GW) leads to  the characteristic
form of vacuum energy density called ``running-vacuum-model''(RVM)-type cosmology~\cite{,JSPRev2013}, in which an (approximately) de Sitter era is
dynamically-induced without the need for ad hoc inflaton fields\,\cite{JSPRev2013,RVMinfl,JSPRev2015,GRF2015,solaentropy}.}

{The upshot is that in our framework a stringy-induced~\footnote{\ntext{In this work, the words `stringy; and `string-inspired' will be interchangeably used to denote a modified version of RVM embedded in string theory, in the sense of using an appropriate gravitational local field theory (see \eqref{sea4}, section \ref{sec:sRVM}, below) viewed as a low-energy limit of strings~\cite{string,olive}. Here `low-energy' means that the energies and momenta of the various fields involved will be much lower than the string mass scale, $M_s$, so that an appropriate derivative expansion in powers of the Regge slope $\alpha^\prime = M_s^{-2}$ (in units $\hbar=c=1$ we work throughout this work) is in operation.}}  RVM form of dynamical vacuum energy emerges naturally and  effortlessly  at this point}. The physical mechanism for the appearance of
GW condensates is the phenomenon known as  `cosmological birefringence' of the GW during inflation\,\cite{stephon} and, as shown in ~~\cite{bms1,bms2,bms3,msb,ms21}, such condensates are responsible for the generation of terms in the vacuum energy density proportional to the fourth power of the Hubble rate  $H^4$. These are precisely the terms that are known to produce inflation  without the need for external inflaton fields, see\,\cite{ms21} for a detailed discussion.   Such a peculiar inflationary phase can be called ``GW-induced stringy RVM inflation'' and may provide a fundamental \textit{raison d'\^etre} for the characteristic mechanism of RVM-inflation which had been studied previously on more phenomenological grounds.
During that GW-induced inflationary era, KR-axion backgrounds varying linearly with cosmic time remain undiluted, leading to eventual matter-antimatter asymmetries (baryogenesis through leptogenesis) in the post-inflationary radiation era.  This is perfectly consistent with the standard thermal history of the universe since during the radiation and matter eras, the gravitational anomalies cancel owing to the generation of chiral {fermionic} matter at the exit phase from the GW-induced stringy RVM inflation, leaving however the due net amount of chiral anomalies which may trigger through non-perturbative effects (e.g. instantons in the Gluon sector of Quantum Chromodynamics part of the matter action) the potentials responsible for the masses of these axions. Ultimately this form of ({massive}) axionic matter can be the substrate of the  {the long sought-for  Dark Matter  in our picture of the cosmic evolution. Our framework, therefore, aims at a possible fundamental explanation for both dark matter and dark energy in our Universe.}

{In our study we also  discuss scenarios for the possible origin of
the primordial GW. We consider a double-step inflationary scenario characterized by a pre-inflationary era
with dynamical broken supergravity (SUGRA), which can be  embedded in string theory. The breaking occurs through
condensation of  the partners of gravitons, or gravitino fields.  This leads to the formation of unstable domain walls, whose collapse in a non-spherically-symmetric manner may trigger the appearance of primordial  GW.}

{Overall, the stringy-RVM emerges in a natural way  during the inflationary stage of our Universe, being its trigger, and provides a fundamental theoretical motivation for the structure of the RVM vacuum energy density,
whose low-energy form has been successfully tested in a variety of
works, and  most particularly as a possible cure for the intriguing  $\CC$CDM tensions\,\cite{tensions,Valentino1,Valentino2}  (namely the troublesome predictions made by the the concordance $\CC$CDM model concerning the  $\sigma_8$ and $H_0$ parameters associated to the structure formation data and the local and CMB measurements of the Hubble parameter, respectively). Remarkably, such tensions can be highly alleviated in the context of the RVM, as shown in the recent analysis\,\cite{EPLtensions}, {which updates and further supports the existing ones\,\cite{rvmpheno1,rvmpheno2,rvmpheno3}. The phenomenological success of the RVM  is certainly a strong spur to continue the theoretical investigations supporting  the RVM cosmology, and the present work is indeed  aligned with this purpose.}}

{The structure of the paper is as follows: in the next section, we describe the basics of the conventional RVM model and discuss its dynamical inflation, without external inflatons. At the same time we consider its string inspired version, which is an attempt to justify the (phenomenologically successful)  RVM structure in the context of the effective string theory framework.  {In Sec.\ref{sec:axionexcit} we clarify the dual
role of the KR axion field both as a ground state vacuum configuration for the string-inspired gravitational
 action and a particle excitation above the vacuum, which plays the role of stiff matter in a pre-RVM inflationary period in the Universe.
 In section \ref{sec:preRVMinl}, we discuss a potential origin of the GW during the pre-RVM-inflationary phase, due to the non-spherical collapse of domain walls, that are created in this phase in scenarios involving the dynamical breaking of supergravities that are embeddable in string theory. Such scenarios lead also to a first hill-top inflationary era in a way we shall explain, together with the necessity for such early inflationary epochs, immediately after the Big-Bang\footnote{{{Except for the cases where we refer to the combined name Big-Bang Nucleosynthesis (BBN), which is a very definite concept, } throughout the paper we use the name Big-Bang, loosely speaking,  as the point where everything starts, somehow the beginning of time. It it exists, it can be conceived, for instance,
as the situation when two brane worlds collide.  It does not mean we have
curvature initial singularities though.}}.  Like the GW-induced second inflation, the first {(hill-top)} inflation also admits an effective RVM
description, entailing no initial singularity, which in the context of the underlying microscopic string theory model may be understood as the consequence of the (infinity of) higher-order space-time curvature terms in the gravitational action}. {In subsection \ref{sec:instsugra} we describe how the universe tunnels to the stable RVM vacuum, as a consequence of instabilities of the dynamically broken supergravity phase. We also make some estimates of the tunnelling rate}.  {Stiff-axion matter becomes subdominant during
the phase of the GW condensation, which leads to the second inflation
of RVM type. In Sec\,\ref{sec:inflscale} we discuss the details of this GW-RVM inflationary phase,
demonstrate the RVM equation of state, using mathematical properties of the gCS anomaly terms, associated with the Cotton tensor, and discuss the scale of the stringy RVM Inflation}. Possible phenomenological implications and outlook
 on the modern-era phenomenology of the RVM are outlined in Sec.\ref{sec:modern}. Finally, conclusions are given in section \ref{sec:concl}}.

\section{Conventional RVM promoted to stringy RVM \label{sec:RVMandsRVM}}

{In this section we review the construction of the stringy RVM for completeness. We commence the discussion by a brief description of the conventional RVM, which will help the reader  understand the connections with the
stringy version, and appreciate which features of the conventional RVM are carried over intact into the stringy case, and which are different.}

\subsection{Conventional RVM}

{The effective form of the vacuum energy density for the RVM reads\,\cite{JSPRev2013,ms21}:
\begin{equation}\label{rLRVM}
\rho^{\Lambda}_{\rm RVM}(H) = \frac{\Lambda(H)}{\kappa^2}=
\frac{3}{\kappa^2}\left(c_0 + \nu H^{2} + \alpha
\frac{H^{4}}{H_{I}^{2}} + \dots \right) \,,
\end{equation}
in which  $\kappa^2 = 8\pi G=\frac{1}{M^2_{\rm Pl}}$ is the four-dimensional gravitational constant, {with} $\MPl = 2.44 \times 10^{18}$~GeV
 the reduced Planck mass, {$c_0 >0$ is a constant which plays the role of
the (current era) de-Sitter cosmological constant,
when \eqref{rLRVM} is considered in the late eras of the Universe, and $H_I$ denotes the Hubble scale at the inflationary epoch.}
{The RVM equation of state is that of an ideal de-Sitter fluid, despite the time dependence of the vacuum energy~\cite{ShapSol1,Fossil2008,ShapSol2,JSPRev2013,JSPRev2015}:
\begin{align}\label{rvmeos}
p^{\rm \Lambda}_{\rm RVM} (H) = - \rho^{\rm \Lambda}_{\rm RVM} (H)~,
\end{align}
where $p^{\rm \Lambda}_{\rm RVM} (H)$ denotes the pressure density.}}
\ntext{It should be stressed that within the initial phenomenological RVM framework, such an equation of state is a {\it postulate}, which the theoretical formalism and the associated Universe evolution are based upon. This RVM {\it equation of state} \eqref{rvmeos} is {\it derived} in the current paper, in the context of the stringy-RVM model of \cite{bms1,bms2,ms21}, in section \ref{sec:eossrvm}. In the context of quantum field theories in curved space-time, non-minimally coupled to gravity, which can also be associated with the RVM framework~\cite{Cristian2020}, a derivation of the equation of state \eqref{rvmeos} will be presented in \cite{Cristian2021}.}

{The coefficients  $\alpha$ and $\nu$ \ntext{in \eqref{rLRVM}} control the running of the vacuum energy density at high and low energies, respectively. Since $\nu$ is more
accessible to the current phenomenological tests, we already know that $|\nu|\ll1$. In fact,
we even know its order of magnitude value from a variety of analysis which performed a fit of the RVM to the overall cosmological data (SNIa+BAO+$H(z)$+LSS+CMB), with the result $\nu=+{\cal O}(10^{-3})$\,\,\cite{rvmpheno1, rvmpheno2,rvmpheno3}, which is reconfirmed  also from the demonstrated ability of the RVM to smooth out the $\CC$CDM tensions, see \,\cite{EPLtensions}.  As for the parameter $\alpha$, it is obvious that it must satisfy $\alpha>0$ in order for the $H^4$ term to possibly trigger inflation,
but its numerical value is fuzzier  since it is connected with the inflationary scale $H_I$ and hence we can only  expect a rough bound on the combination of the two. In the stringy-RVM framework $\alpha$  is found  relatively small, say in the range $0<\alpha\lesssim 0.1$.}

{The generic structure \eqref{rLRVM} was originally motivated from renormalization group arguments, in which $\nu$ acts as the coefficient of the $\beta$-function associated to the running of $\rho^{\Lambda}_{\rm RVM}(H) $ ~\cite{ShapSol1,Fossil2008,ShapSol2}, so not surprisingly $\nu$
is expected to be rather small also on solid theoretical grounds;  for a
review and subsequent developments see ~\cite{JSPRev2013,JSPRev2015} and references therein. However,  it is important to stress that insofar  as we are concerned in the present study  the  $H^4$ terms responsible for RVM inflation are generated as a direct  consequence of the condensation of the anomalous gCS-type interactions of the KR axions with gravity.  Therefore, the crucial terms which trigger inflation are secured in this framework. As for the $H^2$ terms, there are various options, and a particular one will be discussed in the present stringy context as well. However,
 it is fair to say that  their existence as quantum effects in the effective action  of QFT in curved spacetime has also been verified explicitly in a recent work ~\cite{Cristian2020}, which is in itself a remarkable result. No less remarkable is the fact that in such a QFT calculation it has also been shown that \textit{no}  $H^4$-terms are, however,  generated as part of the QFT effective action.  All of the ${\cal O}(H^4)$- terms which are generated involve cosmic time derivatives (e.g.
$\dot{H}^2, H^2\dot{H}, H\ddot{H}$), and hence they all  vanish for $H=$const.   As a result, our stringy RVM inflation provides a genuinely different input, the $H^4$ term,  which characterizes a new type of inflation that is different from the higher-curvature $R^2$-inflation (Starobinsky inflation~\cite{staro}), in which once more only higher order terms which vanish for $H$=const. are generated.  Indeed, recall that in the latter model, inflation is characterized by a phase where the rate of change
of $H$ is constant (i.e. $\dot{H}=$const. rather than
$H=$const.)  For this reason RVM-inflation and  $R^2$- inflation are two fundamentally different scenarios for describing the very early universe, both compatible with the current observations though.  See  \cite{ms21} for a review and  detailed comparison between RVM-inflation and Starobinsky inflation}\footnote{{Let us note that RVM-inflation should not
be identified with just $H^4$-inflation.  It is however true that such a power term is just the canonical or minimal realization enabling  RVM-inflation, and it is the one which is precisely motivated within the stringy-RVM framework under discussion here. However, general RVM-inflation can be triggered by any higher order (even) power  $H^{2n}$ or an arbitrary combination of even powers beyond $H^2$, see e.g. \,\cite{RVMinfl,JSPRev2015,GRF2015,solaentropy} for explicit inflationary solutions of these higher order RVM models. Note that the even character of the power is necessary to
insure that the model is compatible with general covariance.}}.

A key ingredient for the embedding of the above RVM structure  into an string effective theory is  the
presence of the KR axion field.  In four space-time dimensions, the latter is represented by a pseudoscalar massless excitation, the KR axion field $b(x)$. Such an axion field is the only one that may couple to gravitational anomalies, through the effective low-energy string-inspired gravitational action\,\cite{ms21}.
{Our starting point is a four-dimensional string-inspired cosmological framework, based on critical-string low-energy effective action of the graviton,  $g_{\mu\nu}=g_{\nu\mu}$, and antisymmetric tensor (spin-one) Kalb-Ramond (KR) fields,  $B_{\mu\nu}=-B_{\nu\mu}$, of the massless (bosonic) string gravitational multiplet, after compactification to (3+1)-dimensions. Dilaton couplings to curvature are generally present as well
in the low-energy effective action of string theory. However,  in Ref. \cite{bms2} we have proven that it is perfectly consistent to work in a constant dilaton background, and for the sake of simplicity we shall still adhere  to this simplest option for our current presentation --  {see Sec.\,\ref{sec:rvmIIstrings},  though}. { We should also clarify at this point that we do not consider the SUSY (supersymmetric) partners of the bosonic supermultiplet in our effective action (e.g. the gravitino and the SUSY partners of the bosonic axions and dilatons, namely the  axinos and  dilatinos, respectively) since we assume that SUGRA has been broken spontaneously at a very early stage and gives a large mass to all of them, hence they decouple. We shall
come back to this point in the last part of our work (cf. Sec.\,\ref{sec:preRVMinl})}.}

\subsection{Stringy RVM \label{sec:sRVM}}

{In our road to show how the RVM formalism can be derived from  string effective theory, we start from the corresponding effective action in the aforementioned self-consistent special case of constant dilatons, which  is assumed throughout}{:\footnote{{We follow the notations and conventions of \cite{ms21}, that is: {time dominant} signature of metric $(+, -,-,- )$, Riemann Curvature tensor
$R^\lambda_{\,\,\,\,\mu \nu \sigma} = \partial_\nu \, \Gamma^\lambda_{\,\,\mu\sigma} + \Gamma^\rho_{\,\, \mu\sigma} \, \Gamma^\lambda_{\,\, \rho\nu} - (\nu \leftrightarrow \sigma)$, Ricci tensor $R_{\mu\nu} = R^\lambda_{\,\,\,\,\mu \lambda \nu}$, and Ricci scalar $R = R_{\mu\nu}g^{\mu\nu}$.}}
\begin{align}\label{sea4}
S^{\rm eff}_B =&\; \int d^{4}x\sqrt{-g}\Big[ -\dfrac{1}{2\kappa^{2}}\, R + \frac{1}{2}\, \partial_\mu b \, \partial^\mu b +   \sqrt{\frac{2}{3}}\,
\frac{\alpha^\prime}{96 \, \kappa} \, b(x) \, R_{\mu\nu\rho\sigma}\, \widetilde R^{\mu\nu\rho\sigma} + \dots \Big],
\end{align}
where $\alpha^\prime = 1/M_s^2$ is the Regge slope of the string, {with
$M_s$  the string scale},
 and $\widetilde R_{\mu\nu\rho\sigma} = \frac{1}{2} \varepsilon_{\mu\nu\lambda\pi} R_{\,\,\,\,\,\,\,\rho\sigma}^{\lambda\pi}$ is the dual of the
Riemann tensor, with  $\varepsilon_{\mu\nu\rho\sigma} = \sqrt{-g}\,  \epsilon_{\mu\nu\rho\sigma}$ (with $\epsilon^{0123} = +1$, {\emph etc.})
the covariant Levi-Civita tensor density (totally antisymmetric in its indices).}

{The structure  $R\, \widetilde R\equiv R_{\mu\nu\rho\sigma}\, \widetilde
R^{\mu\nu\rho\sigma}$ is the gravitational Chern-Simons (gCS) term, and} {the anomalous CP-violating interaction $b\, R\, \widetilde R$
in \eqref{sea4}} proves essential for the creation of anomaly condensates, induced by GW perturbations, which in turn lead to the generation of terms in
the vacuum energy density proportional to the fourth power of the Hubble rate  $H^4$ ({the core piece of RVM-inflation in its canonical form}).
 {These terms indeed induce inflation} without the need for external ({ad
hoc}) inflaton fields~\cite{JSPRev2013,RVMinfl,JSPRev2015,GRF2015,solaentropy}. In our case, though, there is a scalar  d.o.f.  associated with the condensate. The slow-roll nature of inflation is linked to the slow-roll behaviour of the KR axion.

\ntext{It should be remarked at this point, that in string theory there is, in general, a non-constant dilaton field $\Phi$ present. Setting it to a constant value, which we assumed in our approach~\cite{bms1,bms2,ms21}, and shall adopt here as well, which fixes the string coupling $g_s=\exp(\Phi)$~\cite{string}, implies additional constraints in the classical solutions of the low-energy theory, including the dilaton equation. However, as discussed in some detail in \cite{bms2}, such a constant dilaton solution can be self-consistently imposed, and does not affect the arguments based on the truncated effective field theory \eqref{sea4} (see also relevant but brief discussion in \cite{olive}). 
It must be stressed at this point that, as discussed in detail in \cite{bms1,bms2,ms21} the effective action \eqref{sea4} is dual to the initial low-energy string effective action based on fields  from the massless bosonic gravitational multiplet ({\it i.e.} spin-2 gravitons, spin-0 dilatons, and spin-1 antisymmetric tensor fields). The 
KR axion field $b$ in \eqref{sea4} and anomaly terms $b \, R \widetilde R$ arise as a result of appropriate implementation of a Bianchi identity constraint (via a Lagrange multiplier field, which is identified with the KR axion $b$~\cite{olive,bms1,bms2}) for the field strength ${\mathcal H}_{\mu\nu\rho}$ of the antisymmetric tensor, whose definition \eqref{csterms} includes the Green-Schwarz counterterms required for anomaly cancellation in the extra dimensional space time of the string~\cite{string}. 
The  equations of motion for the dilaton and antisymmetric tensor fields in the effective string action in the initial formalism read (our arguments do not affect the graviton equation of motion, which admits a constant dilaton solution straightforwardly)~\cite{bms2}:
\begin{align}\label{stringeom}
&{\rm dilaton}: \quad \frac{2}{\kappa^2} \nabla^\mu \partial_\mu \Phi - \frac{2}{3} e^{-4\Phi} \mathcal H_{\mu\nu\rho}\, \mathcal H^{\mu\nu\rho} + \frac{\delta V(\Phi)}{\delta \Phi} =0 , \\
&{\rm antisymmetric~tensor}: \quad \nabla^\mu \Big(e^{-4\Phi} \, \mathcal H_{\mu\nu\rho} \Big)= 0 ,
\end{align}
where $V(\Phi)$ is a dilaton potential that can be generated through string loops, and its purpose will be justified below. In four space-time dimensions, the solution of the equation for the antisymmetric tensor field can be satisfied 
by the ansatz: $e^{-4\Phi} \, \mathcal H_{\mu\nu\rho} \propto \varepsilon_{\mu\nu\rho\sigma} \, \partial^\sigma \overline b(x)$, where $\overline b(x)$ is a classical field that plays the r\^ole of a background for the KR axion field of \eqref{sea4}. Upon using this ansatz in the dilaton equation, assuming isotropy and homogeneity for $b=b(t)$, with $t$ the cosmic time, and imposing $\Phi \simeq $ constant, we arrive at: 
\begin{align}\label{phicons}
(\dot{\overline b})^2 = - \frac{1}{2}\, \frac{\delta V(\Phi)}{\delta \Phi} \Big|_{\Phi \simeq {\rm const}} \gtrsim 0. 
\end{align}
A typical  scenario that satisfies
\eqref{phicons}, with $\dot b \simeq {\rm constant}$, of interest to us here ({\it cf.} Eq.~\eqref{eq:bfluxconst}, section \ref{sec:afterinfl}), would be, for instance, that of a `run away' 
pre-big-bang type~\cite{prebigbang} dilaton potential, in the range where the dilaton slowly approaches a constant value asymptotically and with a decaying trend   $\delta V(\Phi)/\delta \Phi <0$. This situation also characterises N=1 globally supersymmetric theories, embeddable in a supergravity/superstring framework~\cite{gs}. The upshot of this discussion is that $\Phi\simeq$~constant appears to be a viable assumption within our framework both as a self-consistent field theory of graviton and KR-axion degrees of freedom and within a generic string inspired approach to RVM. For more discussion we refer the reader to \cite{bms2}, and references therein.}

The KR  axions are characterised by a ``stiff'' equation of state, $w=+1$, and in the scenario of \cite{ms21} dominate the {pre-RVM-inflationary
}epoch. At this stage we mention that primordial stiff-matter dominance in the Universe has been discussed by Zeldovich, but in the context of a phenomenological cold gas of baryons\,\cite{stiff}, see also \cite{stiff2}
for considerations along similar lines. In our case, stiff matter is connected with properties of the KR axion in the early universe, and thus the
physics in our scenario is very different. Moreover, in our case~\cite{ms21} the stiff-matter dominance occurs in a pre-RVM-inflationary era, as already mentioned, {and is linked to fundamental degrees of freedom of the
bosonic string multiplet.}

During the RVM-inflationary era, KR-axion backgrounds of a specific type,
varying linearly with cosmic time,  appear as consistent solutions
of the equations of motion obtained from the effective action. These remain undiluted, and  lead to an eventual matter-antimatter asymmetries (baryogenesis through leptogenesis) in the post-inflationary radiation era in
models involving right-handed neutrinos~\cite{decesare}.

At the exit from inflation, chiral {fermionic} matter is generated, which is itself responsible
for the generation  of {both, chiral and gravitational anomalies. The latter can be arranged to cancel the
primordial ones that lead to inflation~\cite{bms1,bms2}. Indeed, the presence of chiral fermionic matter implies the addition to the
effective action \eqref{sea4}, of fermionic terms of the form (for details, see \cite{bms1,bms2})
\begin{align}\label{fermion}
S_F =  S_{F}^{Free} + \int d^{4}x\sqrt{-g}\, \Big(  \frac{\alpha^\prime}{\kappa}\, \sqrt{\frac{3}{8}} \, \partial_{\mu}b \Big)\, J^{5\mu}    - \dfrac{3\kappa^{2}}{16}\, \int d^{4}x\sqrt{-g}\,J^{5}_{\mu}J^{5\mu}  + \dots \Big] + \dots,
\end{align}
where $S_{F}^{Free}$ denotes free fermion kinetic terms, and
$J^{5\mu} = \sum_{j} \overline \psi_j \, \gamma^\mu \, \gamma^5 \, \psi_j $ is the axial current, with the summation being over appropriate fermion species $\psi_j$ of the matter sector, e.g. charged chiral quarks or leptons in the SM sector, and other chiral fermions in general in string effective theories, including right-handed neutrinos. The coefficients of the KR-axion-axial-fermion-current interaction term in \eqref{fermion} is dictated by the role of the field strength of the KR antisymmetric tensor field as torsion~\cite{bms1}.
At the exit from inflation, all fermion matter is assumed massless and chiral in our model.\footnote{{We stress, however, that massive right-handed neutrinos are crucial for leptogenesis in our scenario~\cite{bms1,decesare}. We do not discuss here the precise way by means of which the right-handed neutrinos acquire their masses. One scenario involves non perturbative string-instanton effects, which can generate shift-symmetry breaking, non-derivative Yukawa-type, interactions of the KR axion $b(x)$ with the right-handed neutrino field~\cite{pilaftsis}, producing radiative masses for the latter
early in the post inflationary era, before leptogenesis.}} }
{The gravitational and chiral-gauge anomaly equation for the chiral fermion current reads:
\begin{eqnarray}
 \label{anom2}
&& \partial_\mu \Big[\sqrt{-g}\, \Big(\sqrt{\frac{3}{8}} \frac{\alpha^\prime}{\kappa}\, J^{5\mu}  -  \sqrt{\frac{3}{8}}\,
\frac{\alpha^\prime}{\kappa} \, \frac{\mathcal N}{192\pi^2} {\mathcal K}^\mu  \Big) \Big] \nonumber \\
&&\!=\!   - \, \sqrt{\frac{3}{8}} \frac{\alpha^\prime}{\kappa}\,  \frac{\mathcal N}{32\pi^2}  \, \sqrt{-g}\,
\Big(e ^2 \, {F}^{\mu\nu}\,  \widetilde{F}_{\mu\nu} + g_s^2 \, {\mathcal G}^{a\,\mu\nu}\,  \widetilde{\mathcal G}^a_{\mu\nu}\Big)
\end{eqnarray}
where $\partial_\mu \Big( \sqrt{-g}\ \mathcal K^\mu \Big) = \sqrt{-g} \, R_{\mu\nu\rho\sigma} \, \widetilde R^{\mu\nu\rho\sigma}$ represents the Chern-Simons gravitational anomaly terms, and
$F_{\mu\nu}$ and $\mathcal G_{\mu\nu}^a, \, a=1, \dots 8$ are the Maxwell and Gluon tensors respectively (with the tilde denoting their corresponding dual in curved space, $\widetilde A_{\mu\nu} \equiv \frac{1}{2} \, \varepsilon_{\mu\nu\alpha\beta} \, A^{\alpha\beta}$, with $\varepsilon_{\mu\nu\alpha\beta}$ the covariant Levi-Civita tensor density), and $e$ and $g_s$ is the electromagnetic and strong couplings, respectively. The quantity $\mathcal N > 1$ denotes  effectively the number of (stringy) chiral degrees of freedom that circulate in the fermion loops of the anomalies, leading to the non conservation of the chiral current \eqref{anom2}. The precise value of $\mathcal N$ depends on the microscopic string theory considered, but it is in general relatively large.}

{In \cite{bms1} and above, for concreteness, we considered $\mathcal N$ to be of the appropriate order so as to cancel the gravitational anomalies in the primordial effective action, with coefficient as set in \eqref{sea4}, that is $\frac{\mathcal N}{192\, \pi^2}  \sim \frac{4}{3 \times 96} $, i.e. $\mathcal N \sim 26$,
as an order of magnitude estimate. The fact that $\mathcal N$ turns out not to be quite an integer is not a problem, since
this can be rectified by modifying slightly the coefficient of the gravitational Green-Schwarz counterterm appearing in the definition of the KR field strength (in differential form notation)
 \begin{align}\label{csterms}
\mathbf{{\mathcal H}} &= \mathbf{d} \mathbf{B} + \xi \,  \frac{\alpha^\prime}{8\, \kappa} \, \Omega_{\rm 3L},  \nonumber \\
\Omega_{\rm 3L} &= \omega^a_{\,\,c} \wedge \mathbf{d} \omega^c_{\,\,a}
+ \frac{2}{3}  \omega^a_{\,\,c} \wedge  \omega^c_{\,\,d} \wedge \omega^d_{\,\,a}, \quad \xi = \frac{3\mathcal N}{8\, \pi^2}~,
\end{align}
with $B_{\mu\nu}=-B_{\nu\mu}$ the antisymmetric tensor field, and $\omega^a_{\,\,\,b\,\mu}$ the gravitational spin connection (we only give here the gravitational parts of the Green-Schwarz counterterms, of relevance to our stringy RVM approach, where only gravitational d.o.f. are considered in the early Universe effective action as external fields). This leads to
appropriate modifications in the corresponding Bianchi identity constraint, which introduces the KR axion $b(x)$ into the effective action as a Lagrange multiplier in the path integral over $\mathcal H_{\mu\nu\rho}$~\cite{bms1}. This is the essence of the gravitational-anomaly cancellation mechanism by chiral matter proposed in \cite{bms1}. This means that in our stringy RVM approach, we replace the gravitational Chern-Simons (gCS) anomalous (counter)term in the effective action \eqref{sea4} by
\begin{align}\label{improvedCS}
S_{\rm gCS} = \int d^{4}x\, \sqrt{-g}\, \xi \,  \sqrt{\frac{2}{3}}\,
\frac{\alpha^\prime}{96 \, \kappa} \, b(x) \, R_{\mu\nu\rho\sigma}\, \widetilde R^{\mu\nu\rho\sigma}, \qquad \xi = \frac{3\mathcal N}{8\, \pi^2}~.
\end{align}
For $\mathcal N=26$ one obtains $\xi =0.99$.
It is this postulate that
leads to the order of magnitude of the estimates regarding the string scale we discussed in \cite{bms1,bms2,ms21} and here.
In general, depending on the microscopic theory considered, one may use a different value of $\mathcal N$ in \eqref{improvedCS}.
In string theory,  one does not expect $\mathcal N$ to be larger than about a 1000~\cite{stephon}. Considering such generic $\mathcal N$,
one can absorb formally $\xi $ in the definition of an effective $\sqrt{\alpha^\prime}_{\rm eff}  \equiv \sqrt{\alpha^\prime}\, \sqrt{\xi}$ in \eqref{improvedCS}. Then
our analysis in \cite{bms2,bms2,ms21} and here is repeated using this effective string scale, which implies larger string mass scales $M_s=1/\sqrt{\alpha^\prime}$, and thus larger corresponding bounds, by a factor $\xi^{1/2} = \sqrt{3\mathcal N /8} \pi^{-1} \lesssim 6$
than the ones considered here (see, e.g. \eqref{msmpl}) and in \cite{bms2}. Thus, our qualitative and quantitative conclusions will remain essentially the same, independently of the precise value of $\mathcal N$.}

{In fact at this point we remark that the generation of chiral matter, which will dominate the Universe at the exit from inflation, implies in any case that globally, at large scales, the universe will recover its FLRW background profile, for which there are no gravitational anomalies.  However, it is the aforementioned precise cancellation of the gravitational anomalies, at the level of the Lagrangian, that ensures their absence at local scales, and thus standard large- and small-scale cosmology after the RVM inflation.
The cancellation of the gravitational anomalies at the end of the RVM inflation is required for consistency of the matter and radiation theory, which should be diffeomorphism invariant at {\it all scales}.\footnote{{This local cancellation will imply the absence of Chern-Simons gravity in our RVM model, and thus the absence of any appropriate signals of parity-violation through local-source-induced gravitational waves, during radiation and matter eras. Thus one should not worry about Big-Bang-Nucleosythesis (BBN) constraints in our context. In the modern era, though, as discussed in \cite{bms1}, where matter becomes again subdominant compared to vacuum energy contributions, gravitational anomalies might resurface.} } However, the chiral anomalies remain \eqref{anom2},
as their presence does not affect the conservation of the stress tensor of the matter and radiation theory, given that the variation of the corresponding Lagrangian terms with respect to the metric tensor vanishes. The presence of chiral anomalies involving the gluon tensor,
may lead to potentials}, and thus masses, for the KR  axions, non-perturbatively generated through, e.g. instanton effects in the gluon sector of Quantum Chromodynamics, which is part of the matter action.
In this way, the KR axion may behave as (a component of) Dark Matter~\cite{bms2}.

\subsection{Undiluted axion background after inflation \label{sec:afterinfl}}

\begin{figure}
\begin{center}
\includegraphics[width=\textwidth]{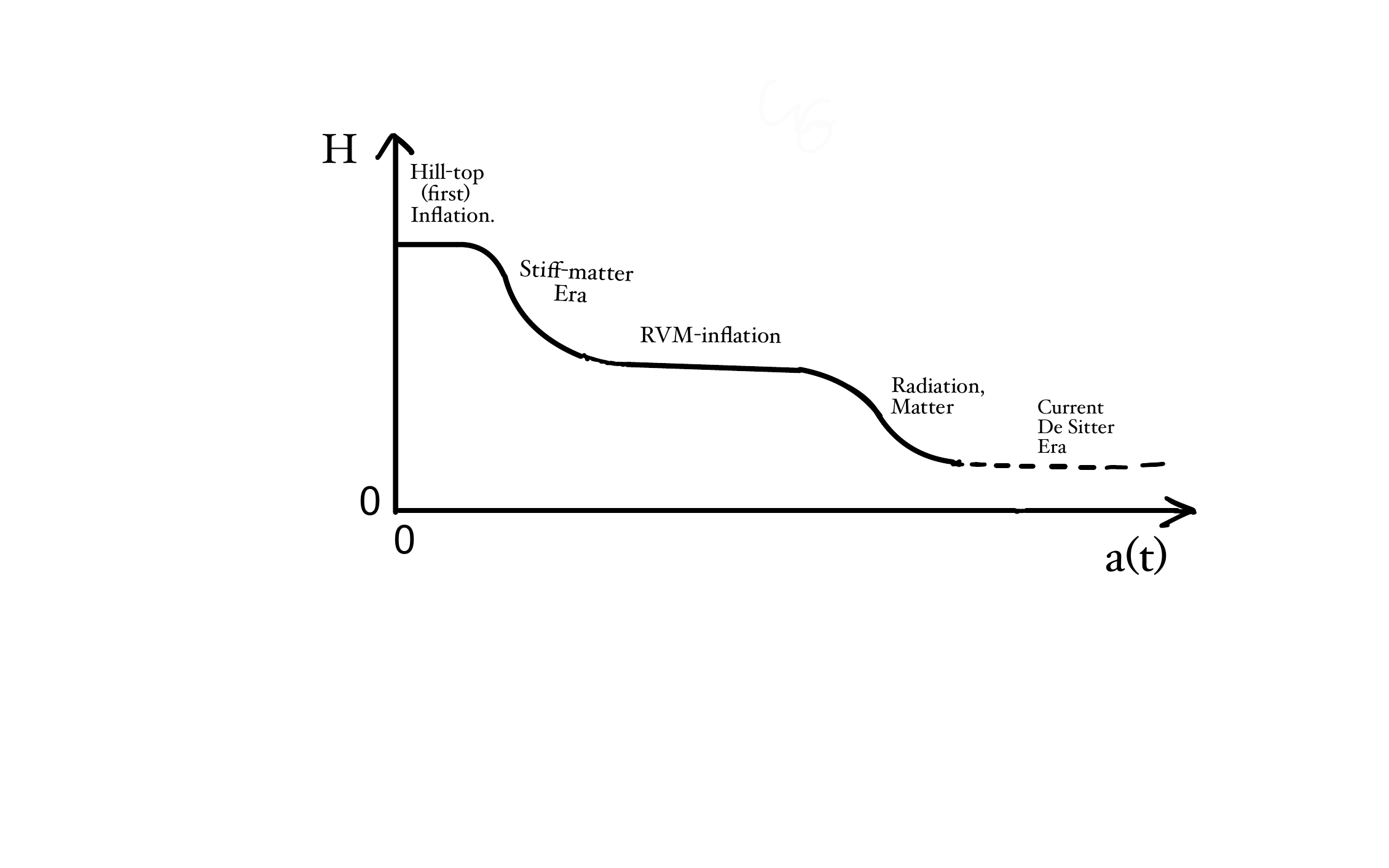}
\end{center}
\vspace{-3.5cm}
\caption{{\it Completion of the RVM inflationary scenario of\,\cite{bms1,bms2,bms3,msb,ms21}.} \it Schematic evolution of the Hubble parameter with the scale factor of an expanding stringy-RVM Universe, {involving two inflationary eras}. The hill-top first inflation, which exists immediately
after the Big-Bang, may be due to dynamical breaking of SUGRA, which characterises the string-inspired Universe at very early stages. At an
effective, low-energy field theory level, the RVM approach also characterises the {first inflation} during the broken SUGRA phase. There
is no initial singularity in an RVM framework, which microscopically in string theory may be attributed to non-perturbative
resummation of higher curvature terms in the string effective action.  {The first SUGRA inflation} {is responsible for washing out any
spatial inhomogeneities in the effective field theory. It is} {followed by the RVM-GW-induced inflation,} {which is the main topic of the approach
of \cite{bms1,bms2}, and the one that can be tested with cosmological data}.  {The two inflationary eras} {are separated by  a peculiar non-RVM phase dominated by excitations of the KR field about its ground state, in combination with the gravitational anomaly, a state that we call here `phantom vacuum', see the text}.}
\label{fig:Hevolrvm}
\end{figure}

{A fundamental result of our previous {analyses\cite{bms1,bms2} (see also~\cite{ms21} for a detailed review and additional considerations)} is that after GW-induced stringy RVM inflation and under appropriate conditions
we are left with an undiluted background of KR axions.
Briefly this comes about as follows.  Owing to the presence of gravitational anomalies,  the anomaly equation {(in the homogeneous and isotropic approximation)}
\begin{equation}\label{pontryagin2}
 \frac{d}{dt}  \Big(\sqrt{-g}\, {\mathcal K}^0 (t) \Big) = \langle \sqrt{-g} \, R_{\mu\nu\rho\sigma}\, \widetilde R^{\mu\nu\rho\sigma} \rangle ~,
 \end{equation}
 becomes nontrivial, meaning that  its \textit{r.h.s} (an average over graviton fluctuations in the inflationary space time perturbed by the primordial GW~\cite{stephon})  does not vanish. Under appropriate conditions this gives a chance to end up with a solution in which $\mathcal {K}^0 (t)
$  (the first component of the anomaly current $\mathcal {K}^\mu (t) $\cite{bms1}) is nonvanishing and constant at the end of inflation: $ {\mathcal K}^0 \simeq$ const.   In this case, considering the Lagrangian field equation for the KR axion field we are left with an undiluted background flux of axion matter at the end of inflation:
\begin{align}\label{eq:bfluxconst}
\dot{\overline{b}}  =  \sqrt{\frac{2}{3}}\, \frac{\alpha^\prime}{96 \, \kappa} \, {\mathcal K}^{0}\simeq {\rm constant}\,,
\end{align}
where the overdot denotes derivative with respect to the cosmic time $t$ in the usual Friedman-Lema\^{\i}tre-Robertson-Walker (FLRW) frame. In stark contrast,  in the absence of gravitational anomalies  there would be no constant ${\dot b}$  ground state configuration  for KR axions at all since the \textit{r.h.s}  of  \eqref{pontryagin2} would vanish identically. As a consequence, the quantity being differentiated in that equation is constant,  and thus $\mathcal{K}^0 (t)\propto \sqrt{-g}\sim a^{-3}(t)$. This result combined with  \eqref{eq:bfluxconst} yields
\begin{align}\label{eq:dotbaminus3}
\dot b _{\rm stiff}\sim a^{-3}(t)\,,
\end{align}
and hence there would be a fast dilution of all the axion matter.  Even though this does not occur in our case at the end of the RVM inflation stage,  it  is actually the precise situation in the stiff matter era, namely  a  pre-inflationary phase (prior to the RVM regime) in the context of the extended picture that we will present in subsequent sections of this work. See the schematic representation of the complete history of the stringy RVM-Universe in Fig.~\ref{fig:Hevolrvm}. {This figure extends and completes our initial picture of RVM inflation presented in \cite{bms1,bms2}.  We will provide more details on its meaning throughout our exposition}.}

{Interestingly, the fact that the KR axion flux after the GW-RVM inflation may
remain approximately constant by virtue of our scenario of  nonvanishing
gravitational anomalies during inflation suggests that we may associate that constant flux with the constancy  of the slow-roll parameter of inflation, $\epsilon$.
It is well-known that  for a period where a generic scalar field drives the cosmological evolution  the rate of change of the Hubble function is related in very good approximation to the rate of change of the scalar field   by $\dot{\varphi}^2 \simeq -2\MPl^2\dot{H}$. (This follows from differentiating Friedman's equation and using the Klein-Gordon equation satisfied by the homogeneous scalar field).   In this way,  the standard slow-roll parameter $\epsilon$, which characterises the inflation period, can
be written as
\begin{equation}\label{eq:SlowRoll}
\epsilon \equiv -\frac{\dot{H}}{H^2}=\frac12\,\frac{\dot{\varphi}^2}{(\MPl H)^2}\ll1\,.
\end{equation}
Obviously, the $\epsilon$ parameter must be small during inflation since
in that period the rate of change of the Hubble function is small.   On  applying the previous equation to the undiluted background of the KR axion feld  $b$ at the end of inflation we are led to
\be\label{dotearly}
\dot{b}=\sqrt{2\epsilon} \, M_{\rm Pl} H\,.
\ee
}
{Because the  RVM is characterised by an approximately constant
Hubble parameter $H \simeq H_I$, (in contrast to Starobinsky inflation), we find a
slow-roll parameter $\epsilon$ similar to the conventional inflaton field (although we stress again that the KR axion is {\it not} an inflaton, as our inflation is realised dynamically):
\begin{align}\label{bfield}
b(t) = \overline b(0) + \sqrt{2\, \epsilon} \, H_I \, t\, M_{\rm Pl}~,
\end{align}
where $\overline b(0)$ is the initial value (at the beginning of RVM inflation)
of the KR axion.  The first (constant) term is essential to generate the GW consensate, whereas the second can be bounded by the condition (see below)  that there is a minimum number of inflationary e-folds during the (approximately)  constant ${\dot b}$  regime. }

{In what follows we will explore important features of the stringy RVM, by first clarifying the role of the KR axion field, which behaves both as matter excitation and vacuum (ground state) contribution. As we shall discuss, during the phase where the primordial GW condense, {the equation of state of all} the vacuum contributions (which carry different signs) conspire to render precisely that of the dynamical vacuum energy in the RVM, i.e.
~\cite{JSPRev2013,JSPRev2015},
\be\label{totvac}
p_{\rm total}^{\rm vacuum} = - \rho_{\rm total}^{\rm vacuum} < 0,
\ee
with the total vacuum energy density being positive ($ \rho_{\rm total}^{\rm vacuum}>0 $).}

\subsection{The Cotton tensor and the deformed Einstein's equations}

{ As discussed in \cite{bms1}}, { there are several contributions to the two sides of \eqref{totvac}, coming from the KR axion field, the anomalous CS terms, associated with the so-called Cotton tensor~\cite{jackiw}, which is traceless, and a GW-induced condensate term,
which behaves as quantum-gravity generated trace of the Cotton tensor. This latter term behaves like a de Sitter cosmological constant term, that dominates the energy density, implying an RVM-like inflationary period, without external inflatons. The dominance {of such condensate over the contribution}
induced by the Cotton tensor is crucial for the overall positivity of the
total vacuum energy density.}

{For completeness, and convenience of the reader, we mention that the
Cotton tensor $C_{\mu\nu}=C_{\nu\mu}$  is constructed from the variation of the
anomalous Chern Simons terms $b \, R \, \widetilde R$ in \eqref{sea4} with respect to the gravitational field $g_{\mu\nu}$:
\begin{align}\label{csgrav}
\delta \Big[ \int d^4x \sqrt{-g} \, b \, R_{\mu\nu\rho\sigma}\, \widetilde R^{\mu\nu\rho\sigma} \Big] = 4 \int d^4x \sqrt{-g} \, {\mathcal C}^{\mu\nu}\, \delta g_{\mu\nu} = - 4 \int d^4x \sqrt{-g} \, {\mathcal C}_{\mu\nu}\, \delta g^{\mu\nu},
\end{align}
where~\cite{jackiw}
\begin{align}\label{cotton}
&{\mathcal C}^{\mu\nu} \equiv  -\frac{1}{2}\, \Big[v_\sigma \, \Big( \varepsilon^{\sigma\mu\alpha\beta} R^\nu_{\, \, \beta;\alpha} +
\varepsilon^{\sigma\nu\alpha\beta} R^\mu_{\, \, \beta;\alpha}\Big) + v_{\sigma\tau} \, \Big(\widetilde R^{\tau\mu\sigma\nu} +
\widetilde R^{\tau\nu\sigma\mu} \Big)\Big]\, , \nonumber \\
&= - \frac{1}{2} \Big[\Big(v_\sigma \, \widetilde R^{\lambda\mu\sigma\nu}\Big)_{;\lambda}  + \, (\mu \leftrightarrow \nu)\Big]\, ,
\quad
v_{\sigma} \equiv \partial_\sigma b = b_{;\sigma}, \,\,v_{\sigma\tau} \equiv  v_{\tau; \sigma} = b_{;\tau;\sigma}.
\end{align}
As follows from properties of the Riemann tensor, the gravitational trace
of the Cotton tensor vanishes~\cite{jackiw}
\begin{align}\label{trace}
g_{\mu\nu}\, \mathcal C^{\mu\nu}= 0~,
\end{align}
and we also have the covariant non-conservation law:
\begin{equation}\label{csder}
{\mathcal C}^{\mu\nu}_{\,\,\,\,\,\,\,;\mu} = -\frac{1}{8} v^\nu \, R^{\alpha\beta\gamma\delta} \, \widetilde R_{\alpha\beta\gamma\delta}.
\end{equation}
}
In the presence of anomalous terms, and the GW-induced condensate, {an effective cosmological constant $\Lambda (H)  \sim\,< \overline b R\, \widetilde R> \ne 0$ is generated}, where
$H$ is the (approximately) constant Hubble parameter during the RVM inflation, and
$b$ denotes the classical solution of the KR axion field equation of motion \eqref{bfield}, the Einstein's equations stemming from the
action \eqref{sea4}, read:
\begin{align}\label{einsteincs}
R^{\mu\nu} - \frac{1}{2}\, g^{\mu\nu} \, R  = \Lambda (H) g^{\mu\nu} + \sqrt{\frac{2}{3}}\,
\frac{\alpha^\prime\, \kappa}{12} \,  {\mathcal C}^{\mu\nu} + \kappa^2 \,
T^{\mu\nu}_{b},
\end{align}
where $T^{\mu\nu}_b$ is the stress tensor of the KR axion-like fields.  {As can be seen, the Cotton tensor generated from the Chern-Simons
term in the action produces a deformation of Einstein's GR\,\cite{jackiw}. Such deformation produces of course a violation of general covariance,  caused by the non-vanishing divergence of the Cotton tensor \eqref{csder} in general spacetimes (in particular , in a spacetime perturbed by a GW),which is a reflex of the gravitational anomaly inherent in that tensor. As a result}, {from \eqref{csder} and the Bianchi identity of the
Einstein tensor, there is an energy exchange between the $b$-system and the
gravitational field, when the anomaly terms are non zero, as happens in the mentioned case of GW perturbations~\cite{bms1}. Once the Universe enters the radiation phase and remains in the FLRW metric the \textit{r.h.s.}
of  \eqref{csder} vanishes and the diff-invariance is restored.  {What is more, for the FLRW universe the Cotton tensor itself vanishes,  ${\mathcal C}_{\mu\nu} =0$, and the deformed Einstein equations \eqref{einsteincs} recover their standard form. }}

{As discussed in \cite{bms1}, the Cotton tensor contributions to the energy density, $C_{00}$,
which are associated with the gravitational anomalies, are negative and overcome the KR axion contributions to the energy density.  It turns out that crucial for the dominance of the condensate de-Sitter-like term in the effective gravitational action, is a transplanckian
value of the KR-field ground state configuration, $\overline b(0)$ at the
onset of the RVM inflation (GW condensation). Such transplanckian values are in agreement with the transplanckian conjecture hypothesis (TCH)~\cite{TCH},  according to which {no modes} with four-momenta of magnitude higher than the Planck scale are allowed to enter in the low-energy
effective
action, thereby promoting the role of the Planck mass as an Ultra-Violet (UV) momentum cutoff scale (cf. Sec.\,\ref{sec:transplank} for further discussion).  The key feature of the string-inspired gravitational action, which describes the evolution and dynamics of the cosmological model we discuss here and in \cite{ms21,bms1,bms2}, is that, despite the transplanckian values of $b(0)$, all the terms in the effective action assume sub-planckian values, thereby respecting the TCH.  {We shall substantiate this
assertion throughout our exposition}.}

\section{Axion matter background and its excitations - the role of stiff axion matter \label{sec:axionexcit}}

{We remark at this stage that the background KR axion field plays the role of a ground state configuration of the underlying string model. On top of this background, there exists a quantum excitation, the KR axion field, which is massless. This massless axion may be thought of as corresponding to the Goldstone Boson of the spontaneously broken Lorentz symmetry
 by the KR background. {The spontaneously Lorentz violating (LV) } background pertains therefore to the {\it vacuum}, and it is in this sense that
we view the approximate configuration
\eqref{bfield} for  $\dot  b $ as pertaining to the RVM. The KR excitations, on the other hand, which are excitations about the vacuum,
constitute ``axion matter" contributions to the energy density {of the Universe}, $\rho^{\rm KR-matter}$. They can dominate an era that preceded the RVM-GW-indiced inflation.
The KR axion matter obeys a stiff equation  of state,
\begin{align}\label{w=1}
 p^{\rm KR~matter} = + \rho^{\rm KR~matter}\, > \, 0\,,
\end{align}
given that the $b$ quantum field does not have a potential during the early epochs of the Universe (such a potential might be generated by non-perturbative instanton effects, but only at the post-RVM inflationary phase,
as discussed in some detail in \cite{bms2}{)}.}

{During this stiff-axion-matter dominance era, prior to GW-induced inflation,
there is obviously no RVM vacuum (cf. Fig. \ref{fig:Hevolrvm}). The Friedman equation in this case reads
\begin{align}\label{stiffdom}
H^2_{\rm stiff} \simeq \frac{\kappa}{3} \rho^{\rm KR~matter} \sim \frac{\kappa}{6} \, ({\dot b})^2\,,
\end{align}
{which via \eqref{eq:dotbaminus3} implies a scaling of the Hubble rate as
follows:}
\begin{align}\label{eq:Hminus3}
H_{\rm stiff} \sim a^{-3}(t)\,.
\end{align}
{It is interesting to check that these results can also be derived
from integrating the local conservation equation for stiff matter, namely
$\dot{\rho}_ {b,\ \rm stiff}+6H\rho_ {b, \rm stiff}=0$, which implies
\begin{align}\label{eq:rhostiff}
\rho_ {b, \rm stiff} \sim a^{-6}\,.
\end{align}
For stiff matter, this energy is just kinetic energy of $b$, whence the relations \eqref{eq:dotbaminus3} and \eqref{eq:Hminus3}:  $H_{\rm stiff}\sim \rho_ {b, \rm stiff}^{1/2}\sim a^{-3}$ .}}

{However, stiff matter becomes subdominant once the GW condense in the gravitational anomalous terms in the action.
We discussed some potential origin of such GW in \cite{ms21},} {and we
shall elaborate further in the next section \ref{sec:preRVMinl}.}

{\section{Potential origins of Gravitational Waves in a pre-RVM-inflationary epoch \label{sec:preRVMinl}}}

In \cite{ms21} we discussed potential origins of the primordial GW, by presenting several pre-RVM-inflationary scenarios for their production.
One of them, involves dynamically broken supergravity (local supersymmetry = SUGRA~\cite{sugra}). In our approach, we consider the simplest SUGRA model, N=1 SUGRA~\cite{sugra}, without gauge fields, which is a prototype model embeddable in string theory, within the spirit of our approach that only gravitational d.o.f.  appear as external fields in the effective gravitational field theory of the very early string universe. Indeed, the gravitinos, which are the spin-3/2 supersymmetric partners of the gravitons in the N=1 SUGRA model, do belong to the gravitational d.o.f.

\subsection{Dynamical Supergravity (SUGRA) Breaking Scenario \label{sec:sugradyn}}

\begin{figure}
\begin{center}
\includegraphics[width=0.9\textwidth]{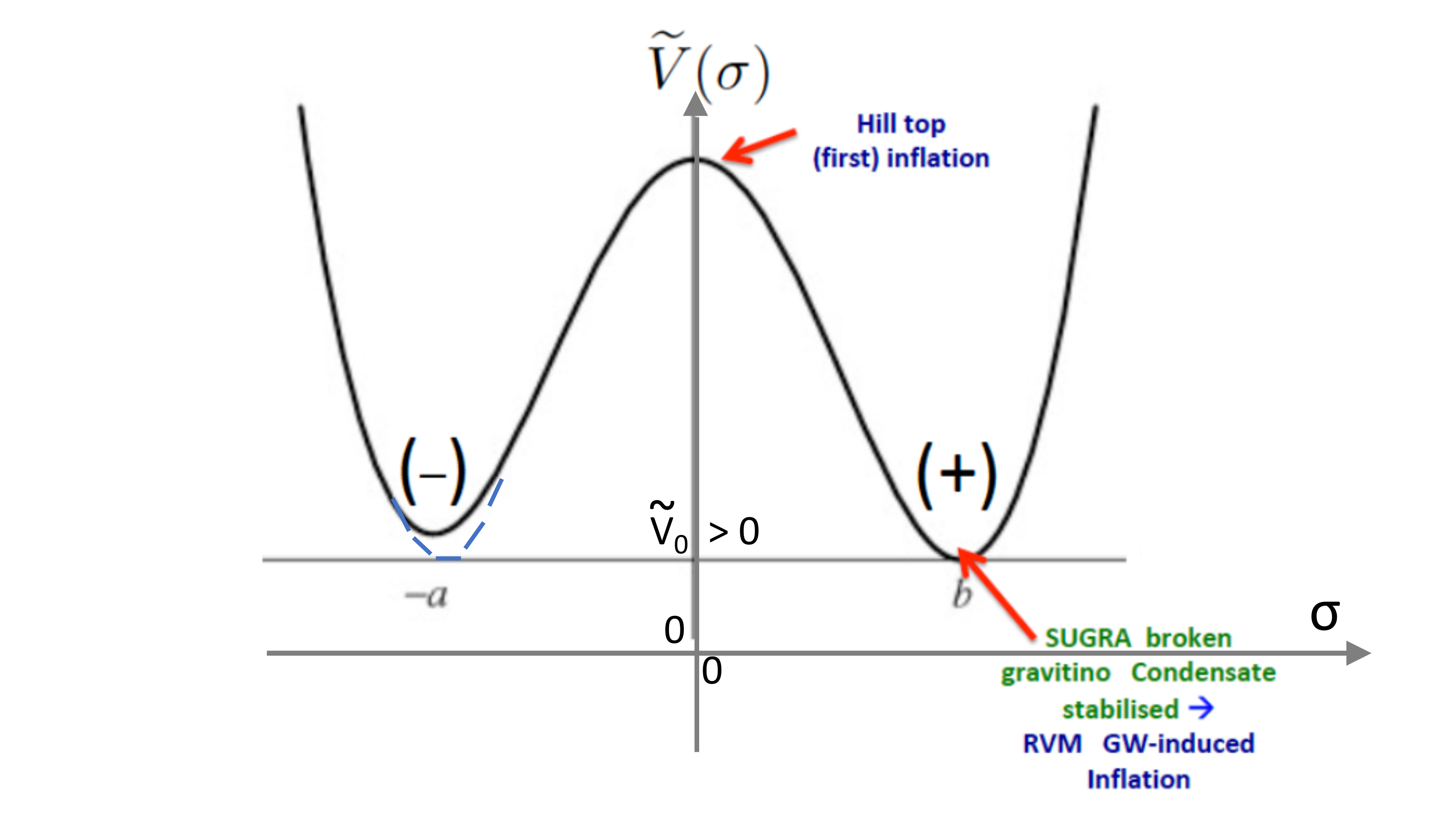}
\end{center}
\vspace{-1cm}
\caption{{\it The real part $\widetilde V(\sigma)$ of the (biased) double-wall potential corresponding to the gravitino condensate $\sigma$ in models with dynamical
breaking of (say $N=1$) supergravities~\cite{houston,ellis}, after percolation effects of the corresponding vacuum bubbles~\cite{ross} are taken
into account ({the figure is not to scale, so the apparent steepness is only an artefact of depiction}). {The bias is one way to lead to unstable domain walls, whose non-spherical collapse produces GW. It is not necessary for the dynamical breaking of SUGRA per se, which can be characterised by a non-biased symmetric double-well potential for the gravitino condensate (dashed lines).}
 The near-zero-field-value region of the potential corresponds to
a `hill-top-inflationary phase, {which corresponds to a
first inflationary phase, preceding the GW-induced RVM inflation~\cite{bms1}}. This latter inflation would occur
at the broken-SUGRA phase, in which the gravitino-condensate field
has been stabilised to a constant {\it translationally-invariant} value ${\sigma_0 = b >0}$ at the bottom $\widetilde V_0 > 0$ of the lower of the two non-trivial vacua.  {Notice that a positive $\widetilde V_0$ is compatible with a broken SUGRA~\cite{houston}}. This plays the role of an one-loop renormalized cosmological constant in the de Sitter scenario of \cite{houston,fradkin}, and its value is found self consistently by minimising the one-loop effective action.} }
\label{fig:bias}
\end{figure}

The breaking of SUGRA occurs through
condensation of gravitino fields. The scalar condensate field $\sigma(x)$ (of mass dimension +2) of gravitinos is characterised by a double-well potential~\cite{houston}, which may be eventually deformed
by `bias' induced due to, say, percolation effects of vacuum bubbles in the effective theory~\cite{ross} (see fig.~\ref{fig:bias}, where the real part of the potential, $\widetilde V$, is depicted as a function of the field $\sigma(x)$). This leads to the formation of unstable
 {domain walls (DW}), whose collapse in a non-spherically-symmetric manner leads to primordial GW.\footnote{{Recall that non-spherical symmetry is necessary here to avoid the implications of  Birkhoff's theorem. Basically, a spherically pulsating or collapsing object cannot emit gravitational waves.}}

There is a non trivial minimum of the double-well potential of  $\sigma(x)$ of  fig.~\ref{fig:bias}, at which the field $\sigma(x)$ is stabilised at its condensate value
defined as $\sigma_c \propto \kappa \langle \overline \psi_\mu\, \psi^\mu \rangle$, where $\propto $ indicate appropriate
numerical coefficients of order one, of no relevance to our qualitative discussion here, and
$\psi_\mu$ is the gravitino Rarita-Schwinger field
of mass dimension 3/2. The condensate is an appropriate mean field solution, independent of space-time coordinates (due to translational invariance of the vacuum), which minimizes the quantum effective potential (or, equivalently, in a Euclidean path integral formulation, the effective action $\Gamma_{\rm eff}$)
\begin{align}\label{loneloopcondens}
 \frac{\delta \Gamma_{\rm eff}}{\delta \sigma }{\Big|_{\sigma = \sigma_c}}=0~.
\end{align}

As discussed in \cite{houston}, condensation of gravitinos results in the latter developing a dynamical mass $m_{\rm dyn} \propto \kappa \sigma_c$, where again the proportionality coefficient is a number of order one. The gravitons on the other hand remain massless, and thus
one has a dynamical breaking of SUGRA.
At this minimum, both, the $\sigma$ field and the gravitino, get a
mass. Depending on the model parameters, this mass can be close to the Planck mass, and thus the gravitino field is integrated  out of the effective field theory, which consists of only massless gravitational degrees of freedom, graviton, dilatons and KR axions. We remark at this point that in the scenarios of \cite{ellis,houston} only $N=1$ SUGRA~\cite{sugra} is considered, as a concrete example of dynamical supersymmetry breaking. In our stringy context, the supersymmetric partners of the dilaton and KR axions would also acquire masses, which, like the gravitino ones,  can be arranged to be close to Planck mass~\cite{houston,sugraRVM},  and thus are also integrated out in the low energy gravitational field theory that describes the RVM inflationary epoch. {This is, of course,  as previously stated, the reason why our effective string theory action was given by Eq.\,\eqref{sea4}, involving only the massless d.o.f. of the bosonic string multiplet (the dilatons are assumed to be constant)}.
Quantum fluctuations about the solution \eqref{loneloopcondens} define the quantum condensate field $\sigma(x)$, which is a fully fledged quantum field.

There might be a hill-top inflation~\cite{ellis} in such scenarios, which takes place at values of the condensate field near zero.
In our context this constitutes a first inflationary phase, preceeding the GW-induced RVM inflation.
In our context, the dynamically-broken SUGRA vacuum is not absolutely stable, in the sense that, as we shall discuss in the next subsection, the one loop SUGRA effective action suffers instabilities, due to imaginary parts. However, from our point of view this is a welcome fact, because it allows tunnelling of the system out of the broken-SUGRA phase and entrance into the (stable) RVM vacuum, which is characterised by its own dynamical inflation. The first inflation ensures that any spatial inhomogeneities are washed out, thus justifying fully the isotropic and homogeneous approximations involved in the study of the stringy RVM inflation of refs.~\cite{bms1,bms2}.

\subsection{Instabilities of the Broken-SUGRA-vacuum and tunnelling to the RVM Vacuum \label{sec:instsugra}}

In \cite{houston} a one-loop approximation was considered when calculating the effective action, which suffices for our purposes due to the weakness of gravity at cosmic times large compared to Planck times, we are interested in for our considerations here (we consider phases of the string Universe after the Big Bang, and hence times larger than the Planck time). The Euclidean formalism was used when computing the path integral of perturbative quantum gravity about de Sitter space time, the de Sitter space time corresponding to the space-time corresponding to the hill-top inflation in our scenario~\cite{ellis}. The supersymmetric formalism~\cite{fradkin} necessitates the introduction of
a bare vacuum energy term of the form
\begin{align}\label{barecc}
\Lambda_0 = - \kappa^2 (f^2 - \sigma^2(x))
\end{align}
where $\sqrt{f}$ the energy scale of supersymmetry breaking. In the absence of the condensate. or when the field is near its zero value, $\Lambda_0 <0$ (anti de Sitter type cosmological constant) as required for unbroken supersymmetry. The renormalised one-loop cosmological constant is positive $\Lambda > 0$, and is determined self consistently by minimising the (Euclidean) quantum effective action~\cite{houston,fradkin}.
This corresponds to the value of the real part of the effective potential  at the (+) minimum, $\widetilde V_0 > 0$, depicted in fig.~\ref{fig:bias}.
Analytic continuation to the Minkowski space time is done only at the very end of the computation.

We note that the effective potential of the gravitino condensate quantum field $\sigma$ has in general imaginary parts, as a result of the (one-loop) graviton contributions~\cite{houston}, which imply terms in the one-loop effective potential proportional to ${\rm ln}(-\Lambda_0)$ ~\cite{houston}, as we review briefly below, for completeness.
By expanding the metric about a de Sitter space-time, which in our context represents the result of the hill-top inflation,
and integrating out graviton (and gravitino in the SUGRA case) degrees of freedom, one arrives at quantum corrected effective actions $\Gamma_{\rm eff}$, in which the de Sitter solution of the space time arises self consistently by looking at minimization of $\Gamma_{\rm eff}$ with respect to the quantum corrected (``renormalised'') cosmological constant $\Lambda$:
\begin{align}\label{loneloop}
 \frac{\delta \Gamma_{\rm eff}}{\delta \Lambda }=0~.
\end{align}
This semi-classical computation can allow for some estimates of the imaginary parts and the tunnelling rate of the false vacuum of the early Universe. The {calculations are performed}  in a Euclidean space-time, where the path integral is well defined,  and at the end one performs analytic continuation  back to Minkowski spacetime.

{ Given that integration of the (massive) gravitino degrees of freedom does not yield imaginary parts, we omit them from the subsequent discussion, referring the interested reader to ref.~\cite{houston} for details.
The one-loop effective action (after performing the graviton integration in a (gauge fixed) {\it Euclidean} quantum-gravity path integral) assumes the generic form~\cite{houston, fradkin, sugraRVM}
\begin{align}\label{effpot}
\Gamma_{\rm eff}^{\rm 1-loop-graviton} = -\frac{1}{2\, \kappa^2} \,  \int d^4 x \sqrt{\widehat g} \Big( \widehat R  - 2  \, \Lambda_1
 + \alpha_1 \, \widehat R + \alpha_2 \, \widehat R^2 + \dots \Big)
\end{align}
where hatted quantities denote background space-time, which can be taken to be the FLRW space time in modern eras, which again enters a de Sitter phase, as appropriate for the computation leading to  \eqref{effpot}~\cite{houston}.
In cosmological, approximately de Sitter, ({\it Euclidean}) backgrounds,
$$\widehat R_{\lambda\mu\nu\rho} = \frac{\Lambda}{3} \, \Big(\, \widehat g_{\lambda\nu} \, \widehat g_{\mu\rho} - \widehat g_{\lambda\rho} \, \widehat g_{\mu\nu}\Big)~,$$
with $\Lambda = {\rm constant} > 0 $ the quantum-corrections-induced cosmological constant, one has the correspondence~\cite{houston,sugraRVM}
\begin{align}\label{corr1}
\int d^4 x \sqrt{\widehat g} =  \kappa^2 \, \frac{24\pi^2}{\Lambda}, \quad
\widehat R \sim 12 H^2,  \quad \Lambda \sim 3H^2,
\end{align}
with $H$ the approximately constant Hubble parameter of the (approximately) de Sitter space-time~\cite{sugraRVM}.}

{In \eqref{effpot}, the quantity
\begin{align}\label{L1}
\Lambda_1 \equiv \Lambda_0 - \kappa^{-2} \, \alpha_0
\end{align}
is the one-loop corrected cosmological constant~\cite{houston,sugraRVM}, after integrating out graviton and gravitino degrees of freedom.}
The coefficients $\alpha_i, i=0,1,2$ in \eqref{effpot} arise from integrating out graviton and gravitino (in N=1 SUGRA models) fluctuations around a de Sitter background at one-loop order~\cite{houston,sugraRVM}. They involve terms with logarithmic dependence on the one-loop renormalised {cosmological constant $\Lambda \sim 3H^2$}. The coefficients arising from integrating out solely graviton (bosonic - superscript $B$)  d.o.f. of interest to us in our subsequent discussion on imaginary parts, look like~\cite{houston,sugraRVM}
\begin{align}\label{coeff}
\alpha_0 = \alpha_0^B &= \kappa^4 \, \Lambda_0^2 \, \Big[\, 0.027 - 0.018 \, {\rm ln} \Big(-\frac{3\Lambda_0}{2\, \mu^2}\,\Big)\,\Big], \nonumber \\
\alpha_1 = \alpha_1^B &= \frac{\kappa^2}{2} \Big[-0.083\, \Lambda_0 +
0.018 \, \Lambda_0 \, {\rm ln}\Big(\frac{\Lambda}{3\mu^2}\Big) + 0.049 \Lambda_0 \, {\rm ln}\Big(-\frac{3\Lambda_0}{\mu^2}\Big) \Big],
\nonumber \\
\alpha_2 = \alpha_2^B &= \frac{\kappa^2}{8} \Big[ 0.020\, + 0.021 \, {\rm ln}\Big(\frac{\Lambda}{3\mu^2}\Big) - 0.014 \,
{\rm ln}\Big(-\frac{6\Lambda_0}{\mu^2}\Big)\Big],
\end{align}
 In the above expressions,
$\mu$ is a renormalization-group (RG) scale of mass dimension +1. It is linked to a proper time cut-off~\cite{fradkin,houston}, which implies that the limit $\mu \to \infty$ ($\mu \to 0$) corresponds to the infrared (UV) limit. In \cite{houston} it was demonstrated that dynamical SUGRA breaking occurs at large (close Planck scale) $\mu$, hence dynamical gravitino mass generation, and thus dynamical breaking of SUGRA,  is
an {\it infrared} phenomenon. From the above considerations, it follows that the ground state assumes an RVM like form~\cite{sugraRVM}, augmented with (mild) logarithmic
dependences ($\sim {\rm ln}H^2$) of the RVM coefficients of the vacuum energy density \eqref{rLRVM} on the one-loop cosmological constant/Hubble parameter  $\Lambda \sim 3H^2$.

We now remark that in \cite{houston}, the gravitino condensate $\sigma_c$, which minimizes the effective potential,  was restricted
to values $\sigma_c^2 < f^2$, which guarantee a real value of the potential at the (+) minimum of fig.~\ref{fig:bias}. However, the quantum fluctuations of the $\sigma(x)$ field remain unrestricted, leading to the aforementioned imaginary contributions, which we stress again, are due to one-loop graviton effects, and thus constitute a pure quantum-gravity effect.

From \eqref{coeff}, it becomes evident that the imaginary terms in the quantum (one-loop) effective action \eqref{effpot} arise from large fluctuations of the quantum field $\sigma(x)$ such that $\sigma(x) > |f| \ne 0$).
The existence of imaginary parts in the effective potential imply an unstable vacuum (+) (and also ($-$)) in fig.~\ref{fig:bias}.
The life time of the false (+) and ($-$) vacua, depends on the parameters of the model, such as $f$, the value of the condensate $\sigma_c$ etc.Its precise value or order of magnitude is of no importance for our study here, except the fact that it is finite.
This is crucial for our scenario, since, as we shall discuss below,  in that case the system
will eventually tunnel through to the (stable) RVM vacuum, after condensation of GW produced by the collapse of unstable DW formed when the gravitino potential is biased somehow.

Let us, therefore, first proceed with estimating the imaginary parts and then the associated decay rate of the false N=1 SUGRA vacuum.
Taking into account the standard Euler's result  that ${\rm ln}(-1) = i\, \pi $, we then obtain from \eqref{effpot},\eqref{corr1},\eqref{L1} and \eqref{coeff} that the imaginary part (``Im'') of the one-loop {\it Euclidean (E)} effective action of $N=1$ SUGRA $\Gamma_{\rm eff}^{(1)\, E}$ \eqref{effpot}, after integrating graviton and gravitino d.o.f. (recalling that the gravitino integration does not lead to imaginary parts~\cite{houston}) read:
\begin{align}\label{impart}
{\rm Im} \Gamma_{\rm eff}^{(1)\, E} \simeq \kappa^2 \, \pi^3 \, \Big( 0.4 \, \frac{\Lambda_0^2}{\Lambda} - 1.2 \, \Lambda_0 + 1.3 \, \Lambda \Big)
\end{align}
with $\Lambda_0$ given in \eqref{barecc}, and $\Lambda \sim 3H^2$, with $H = H^{\rm first}_I$ the (approximately) constant Hubble parameter of the (approximately) de Sitter space-time corresponding to the first hill-top inflation~\cite{ellis}. In our scenarios, as explained above (see fig.~\ref{fig:Hevolrvm}), $H^{\rm first}_I  > H_I^{\rm second GW-RVM} $, where $H_I^{\rm second GW-RVM}$ is the scale of the second inflation, of RVM type, induced by GW, which occurs at an era that succeeds the SUGRA breaking era that  is the focus of our discussion in this section~\cite{bms1,bms2,ms21}.

The reader should notice that the (one-loop) quantum-gravity nature of the instability in the broken SUGRA vacuum is reflected in the presence of $\kappa^2$ factors on the right-hand side of  \eqref{impart}. As already mentioned, we assume here that the hill-top inflation occurs at scales below the Planck scale, so that gravity may be assumed weak and the above estimate is reliable. Otherwise, the full  microscopic string theory needs to be involved in the calculation, which at present prevents us from making any estimates on the  decay rate of the false vacuum. We hope though that our arguments based on one-loop quantum gravity will not change the qualitative nature of our conclusions, namely that there exist a finite decay rate of the false vacuum, and thus a finite life time of the inflationary phase, which is expected to be of order $\sim 1/H$, with $H$ the Hubble parameter of the inflationary epoch under consideration (the life time of the tunnelling process \eqref{lifetime} could be much longer than the duration of the inflationary era of course, that is the period of constant $H$, but for consistency, of course, it cannot be shorter than that !).

To give a rough estimate of the decay rate of the false SUGRA vacuum in our case, we should adopt the generic formalism for calculating decay rates of false vacua~\cite{falsevacua} in field theoretic systems in the absence of gravity, but bearing in mind the important caveat of course that as here we deal with gravitational systems, and hence the gauge invariance of the associated rate of decay of false vacua is an important topic that needs to be addressed carefully. Such a task falls way beyond our scope in the present article, but we need to state it for the benefit of the reader, so as to be aware of the limitations of our claims here.

From the relevant literature~\cite{falsevacua} it seems there is a rather general consensus pointing towards the following formula for the (gauge invariant) decay rate per unit volume $V$, $\gamma$,  of unstable vacua in field theory, based on a one-loop approximation:
\begin{align}\label{decayrate}
\frac{\Gamma_{\rm false}} {V} \equiv \gamma & \sim \lim_{V,T^E \to \infty} \, \frac{2}{V\,T^E}\, \Big| {\rm Im} \exp(-\Gamma_{\rm eff}^{(1)}[\phi_c^E] ) \Big|\nonumber \\
& =  \lim_{V,T^E \to \infty} \, \frac{2}{V\,T^E}\, \exp\Big(-{\rm Re}(\Gamma_{\rm eff}^{(1)}[\phi_c^E])\Big) \, {\rm sin}\Big({\rm Im}(\Gamma_{\rm eff}^{(1)}[\phi_c^E]) \Big)
\end{align}
where $\Gamma_{\rm false} $ is the decay width which is the inverse life-time of the false vacuum,
\begin{align}\label{lifetime}
\tau_{\rm false} = \frac{\hbar}{\Gamma_{\rm false}} ~,
\end{align}
and  $\Gamma_{\rm eff}^{(1)}[\phi_c^E] $ is the Euclidean one-loop effective action, evaluated at a stationary point $\phi^E=\phi_c^E$, such that $\delta \Gamma_{\rm eff}^{(1)}[\phi^E] /(\delta \phi^E) \Big|_{\phi^E=\phi_c^E}=0$. The thermodynamic limit is taken in \eqref{decayrate}
in which the four volume $V T^E$ (where $V$ is the three volume, and $T^E$ is the euclidean time interval)  is taken to infinity.

In our curved-space-time situation, the real and imaginary parts of the one-loop effective action of the N=1 SUGRA
have been computed in \cite{houston} in a fixed gauge, and the role of the $\phi_c^E$ is played by the gravitino condensate. As we have mentioned above, to prove the gauge invariance of the corresponding decay rate in the gravitational case is non trivial and will not be attempted here. We only mention that we may provide an estimate of the
corresponding decay rate of the false N=1 SUGRA hill-top inflation vacuum by using \eqref{decayrate} but
replacing the Euclidean four-volume volume $VT^E$ by the de Sitter volume
\eqref{corr1}, $\kappa^2 \, \frac{24\pi^2}{\Lambda}$. The thermodynamic limit would correspond to the limit where the
one-loop cosmological constant $\Lambda \to 0$, a limit that has been considered in the first paper of ref.~\cite{houston}.
In the limit $\Lambda \to 0^+$ the one-loop-corrected effective action \eqref{effpot} is dominated by weak quantum-gravity corrections ($\propto \kappa^2$)~\cite{houston}, and hence, the decay rate \eqref{decayrate} can be approximated to leading order in $\kappa^2$:
 \begin{align}\label{L0}
\gamma_{\rm \Lambda \to 0}  \sim \frac{\pi}{30} \Lambda_0^2 [\sigma_c] = \frac{\pi}{30} \kappa^4 (f^2 - \sigma_c^2)^2~,
\end{align}
using \eqref{barecc}, and assuming a situation where $\kappa^4 (\sigma_c^2 - f^2) \ll 1$~\cite{sugraRVM}.

However, in the context of our hill-top inflationary scenario, $\Lambda$ should be kept finite, of order ({\it cf.} \eqref{corr1})  of the squared of the Hubble parameter during this first inflationary era,
\begin{align}\label{lHI}
\Lambda \sim 3 \, (H^{\rm first}_I )^2.
\end{align}
The larger the volume (in units of Planck volume), the closer the approximation to the field-theoretic thermodynamic limit.
In that case, the decay rate
\eqref{decayrate} leads to a more complicated, but still {\it finite} expression, where the entire one-loop effective action as computed in \cite{houston} is required, and of course the fact that the tunnelling life time should be at least of order of the duration of the inflation $1/H_I$, i.e. the period in which $H_I$ is approximately constant. These requirements will determine the tunnelling time towards the RVM vacuum.  As a concrete example of this more general case, we may assume a situation where
\begin{align}\label{L0to0}
\Lambda_0 [\sigma_c]\to 0^-,
\end{align}
that is a case where $\sigma_c \sim f$. This limit has to be understood as a shorthand notation of the situation in which $|\Lambda_0|$
is smaller than any other scxale in the problem, but {\it finite}, never reaching zero.
 We also assume that
\begin{align}\label{fsigma}
\sigma_c \kappa^2 \ll 1,
\end{align}
just for calculational convenience.
In such a case, the results of \cite{houston,sugraRVM} imply that the real part of the de Sitter N=1 SUGRA Euclidean effective action \eqref{effpot}  is dominated by the terms
\begin{align}
{\rm lim}_{|\Lambda_0| \ll \Lambda, \,\kappa^2 \sigma_c \ll 1}{\rm Re}\Big(\Gamma_{\rm eff}^{(1)E} \Big) \simeq \frac{24\pi^2}{\Lambda} \alpha_2  \, 144 (H^{\rm first}_I )^4 \sim
-16\, \pi^2 \, {\rm ln}(\frac{|\Lambda_0|}{\mu^2}) \, \kappa^2 \, (H^{\rm first}_I )^2
\end{align}
where we used \eqref{lHI} and
\begin{align}\label{a2}
\alpha_2 &= \alpha_2^B + \alpha_2^F \simeq \kappa^2 \Big( 0.020 + 0.021 \, {\rm ln}(\frac{\Lambda}{3\mu^2}) - 0.014 \, {\rm ln}(-\frac{\Lambda_0}{\mu^2}) \Big) \nonumber \\
& + \kappa^2 \Big( 0.029 + 0.014 \, {\rm ln}(\frac{\kappa^2 \, \sigma_c^2}{\mu^2}) - 0.029 \, {\rm ln}(\frac{\Lambda}{\mu^2}) \Big)
\stackrel{\Lambda_0 \to 0} {\simeq} - 0.014\, \kappa^2 \, {\rm ln}(\frac{|\Lambda_0|}{\mu^2}),
\end{align}
which contains the combined contributions from the one-loop integration of gravitons (B) and (massive) gravitinos (F) in the special limit \eqref{L0to0}, \eqref{fsigma} we are considering here~\cite{houston,sugraRVM}.
In the above relations, as already mentioned, the limit $|\Lambda_0| \to 0$ is understood as implying a small but finite quantity, as compared to other scales in the problem, such as $\mu$, $\sigma_c$, $\kappa^{-1}$ and $\Lambda \sim H^{\rm first}_I$.
Moreover, as discussed in \cite{houston,sugraRVM} and mentioned previously, the renormalization-group scale $\mu$ can be taken to be close to the (reduced) Planck mass scale, given that dynamical breaking of SUGRA occurs in the infrared region of $\mu$ (the reader should recall that $\mu$ is related to a proper time cutoff~\cite{fradkin}, hence this peculiarity that $\mu \to \infty$ is related to the infrared limit).

On the other hand, from \eqref{impart}, we have for the imaginary parts of the effective action
\begin{align}\label{impart2}
{\rm lim}_{|\Lambda_0| \ll \Lambda\, \kappa^2 \sigma_c \ll 1} {\rm Im} \Gamma_{\rm eff}^{(1)\, E} \simeq \, 3. 9 \, \pi^3 \,  \kappa^2  \, (H^{\rm first}_I )^2
\end{align}
In the approximation $(H^{\rm first}_I )^2 \kappa^2\ll 1$, which might be a valid one in case the first inflation scale is only a couple of orders smaller than the reduced Planck scale $\kappa^{-1}$, with the RVM inflation scale being even lower (at least some five orders of magnitude smaller than
$\kappa^{-1}$, as current experiments indicate, {\it cf.}  \eqref{rstar} below).  In such a case the decay rate \eqref{decayrate} of the false SUGRA vacuum can be approximated by
\begin{align}\label{decayrate2}
\gamma & \simeq \lim_{(H^{\rm first}_I )^2 \kappa^2 \gg1 } \, \Big[\frac{(H^{\rm first}_I )^2 }{4\, \pi^2\, \kappa^2}\, \Big|\frac{\mu^2}{|\Lambda_0|}\Big|^{-16\pi^2\, (H^{\rm first}_I )^2 \kappa^2} \,  3. 9 \, \pi^3 \,(H^{\rm first}_I )^2 \kappa^2 \Big]\nonumber \\
& \simeq  \pi \, (H^{\rm first}_I )^4 \, \Big|\frac{\mu^2}{|\Lambda_0|}\Big|^{-16\pi^2\, (H^{\rm first}_I )^2 \kappa^2}, \qquad \kappa^2\,(H^{\rm first}_I )^2  \ll 1 , \quad \, |\Lambda_0| \ll \mu^2~.
\end{align}
Notice that this result is consistent as it leads to life times for the false vacuum larger than the inverse inflationary scale $1/H^{\rm first}_I $, since $\mu^2 \gg |\Lambda_0|$.  If we take, as a concrete case, $\mu^2 \kappa^2 ={\mathcal O}(1)$, then $\mu^{-2} |\Lambda_0| \sim \kappa^2 |\Lambda_0| \ll 1$, and that
the factor $ \Big|\frac{\mu^2}{|\Lambda_0|}\Big|^{16\pi^2\, (H^{\rm first}_I )^2 \kappa^2}$ in \eqref{decayrate2} is of order $\mathcal O(1)$, implying that the decay rate $\Gamma_{\rm false}$ \eqref{decayrate} is of the order of the hill-top inflation scale $H^{\rm first}_I$, and thus the life time \eqref{lifetime} of the order of duration inflation, $1/H^{\rm first}_I$ .

{\subsection{Percolation bias on the Gravitino Potential, DW formation  and KR axions \label{sec:statbias}}}

Let us discuss now how one can introduce a bias in the effective potential which will result in unstable DW formation and thus GW production due to non spherical collapse of the former.
One way of introducing a bias in the potential is through the percolation effects suggested in \cite{ross}, which we follow here as a specific example. For other ways of forming GW we refer the reader to ref.~\cite{ms21}. The percolation effects imply that the occupation probabilities of the two vacua (+), ($-$) are different, introducing a statistical bias. We shall not repeat the computation  here, but we refer the interested reader to \cite{ross} and, for the specific case of the stringy RVM, to \cite{ms21}. We only stress that the percolation implies a biased non-equilibrium phase transition, which leads to instabilities of the formed domain walls, whose non-spherical collapse leads in turn to GW that we need in our scenario for obtaining an RVM inflationary phase, upon GW condensation~\cite{bms1,bms2}.

If one considers the one-loop double-well-shaped gravitino potential discussed above, whose real part $\widetilde V$ is plotted  in fig.~\ref{fig:bias}, in the context of a FLRW background space-time, with Hubble parameter $H$, then the  equation of motion for the (homogeneous and isotropic) gravitino condensate field $\sigma$  is
\be\label{potgrad}
\ddot \sigma + 3 \, H \, \dot \sigma = \frac{\partial \widetilde V}{\partial \, \sigma}.
\ee
As discussed previously, the potential can generate a hill-top ``first'' inflation near the origin
$\sigma=0$~\cite{ellis}, which is assumed to take place
in the very early epoch after the Big-Bang and precedes the GW-condensate-induced RVM inflation of \cite{bms1}, which constitutes the inflationary era to be matched to observations. We stress once again that, by
assuming such an early first inflation, one ensures that at its end, any spatial inhomogeneities of the gravitino-condensate scalar field $\sigma$
have been washed out, and hence \eqref{potgrad} is valid to an excellent approximation.
It goes without saying that we assume here that the {gravitino condensation}
phase takes place after the string-dominated phase of the Universe, near the Big-Bang, where  higher curvature and purely string effects are in operation and might be responsible for the absence of any initial-singularity in the Big-Bang Universe~\cite{art}.
Thus, the gravitino condensation phase  is an intermediate phase between the Big-Bang and the
GW-induced-RVM inflationary phase. The duration of such epochs depends on
the microscopic string theory details.

As explained in \cite{ms21}, the gravitino condensation process is  a {\it
non-thermal equilibrium} phase transition, given that only gravitational interactions are involved in the formation of the condensate, specifically
the four-gravitino interactions~\cite{houston}, as a result of the inherent fermionic `torsion' terms of SUGRA models~\cite{sugra}.
We reiterate once more, that the scale of the breaking of local supersymmetry via the gravitino-condensate formation is taken to be much higher than the inflationary scale of the second RVM GW-induced inflation~\cite{bms1} (see fig.~\ref{fig:Hevolrvm}). The latter occurs at a subsequent phase of this Universe evolution, after GW are formed due to the collapse of unstable  DW. This implies that the scale of the first
inflation is also much higher, close to Planck scale. The tunnelling rate of the unstable dynamically broken SUGRA vacuum could be comparable in order of magnitude to the duration $\tau_I^{\rm first} $ of  the hill top inflation (that is the phase where $H^{\rm first}_I $ is constant,
$\tau_I^{\rm first} \sim 1/H^{\rm first}_I $), as seen, for instance, in some concrete examples discussed in the previous subsection \ref{sec:instsugra}.

The formation and collapse of DW, and eventual condensation of GW, then, is considered as taking place during this tunnelling process. It should be clarified that despite the instability of the dynamical SUGRA phase, the massive nature of the gravitinos remains, and thus they, along with the massive condensate, can be safely integrated out, so that the theory passes to the gravitational theory of \cite{bms1,bms2,ms21} consisting of massless d.o.f. of the string gravitational multiplet (graviton, KR axions and (constant) dilatons).

In this respect, we may also assume that, after the hill-top first inflation, during the decay of the gravitino, stringy KR axions are produced (assuming that the model is  viewed as an effective field theory embedded appropriately in string models). {As mentioned previously}, the latter, being massless, without a potential, are characterised {by a stiff} equation of state: $w=+1$, which implies  a scaling of the Hubble parameter {of the form
\eqref{rvmeos}, which we can write as}:
\be\label{Hstiff}
H_{\rm stiff~axion} = \frac{H^{\rm first}_I }{a(t)^3} ~,
\ee
where we normalise \eqref{Hstiff} such that $a(t_i)=1$, where $t_i$ is the cosmic time corresponding to the exit from the first hill-top inflation.  {Recall that for ordinary (pressureless) matter the dilution law is softer than for stiff matter: $H_{\rm matter}\sim a^{-3/2}$.}
Since this `first' inflation does not lead to observable effects in the CMB~\cite{Planck}, given the existence of the second RVM-like GW-induced inflationary phase, we need not worry about fine tuning the parameters to ensure the right slow-roll phenomenology, and hence, as already mentioned, its scale, $H_i$, could be easily assumed as lying much higher than that of the second GW-induced RVM-inflation, that is, close to Planck energy scale.

{In view of the scaling \eqref{stiffdom} of the energy density of the stiff KR matter, which leads to \eqref{Hstiff}, the latter will eventually cease to be dominant, giving way to GW, coming from the collapse of  DW,  as discussed above. The GW eventually condense, leading to non-trivial
gravitational anomalous terms, coupled to the ground state of the KR axion field (the KR background), which leads to RVM inflation as discussed in
\cite{bms1,bms2}, and reviewed above.}
The formation of GW from the collapse of DW  leads to the GW-induced RVM inflationary epoch, whose constant  Hubble parameter
can be matched smoothly with \eqref{Hstiff} at the onset of the RVM inflation, as shown in Fig.~\ref{fig:Hevolrvm}.  {As we can see from that figure, there is a steep descend of $H$ following the law \eqref{Hstiff} in the stiff matter pre-inflationary era.}  The figure depicts a schematic evolution of the stringy RVM universe with two inflationary eras before
the current ({approximately}) de Sitter era.

Since the stiff-axion-dominated phase occurs for very early epochs of the
Universe, we may easily assume
that the potential-gradient term on the right-hand-side of \eqref{potgrad} can be ommitted when compared to the gravitational friction term $H \, \dot \sigma = H_{\rm stiff~axion} \, \dot \sigma \gg \partial \widetilde V/\partial \sigma$.
Such a condition also characterises the exit from the first hill-top inflationary phase.
In this case, to leading order, the solution of \eqref{potgrad} is an approximate constant {\it classical} gravitino condensate field $\sigma_{\rm
cl}$:
\be\label{constcond}
\sigma_{\rm cl}  \simeq \vartheta = {\rm constant}
\ee
where $\vartheta$ is essentially arbitrary (see also discussion in \cite{ross}). We note that among the allowed constants $\vartheta$ are of course the vacuum-expectation-values (VEV) of the condensate, for which $\partial \widetilde V/\partial \sigma =0$~\cite{houston,ellis}.

This (approximate) constancy of the classical part of the condensate field, $\sigma_{\rm cl}$ persists right up to the first inflationary phase, as one goes backwards in cosmic time. The importance of this constant solution in inducing
statistical bias ({inequality}) in the occupation probabilities of the two vacua of Fig.~\ref{fig:bias} is explained in detail in \cite{ms21}, where we
also discuss the formal necessity of the first inflation in allowing us to perform concrete computations for the percolation probability and thus understand better the evolution of the DW network in this early phase of the string-inspired Universe. This first hill-top inflationary era
also
implies, as already mentioned, that any spatial inhomogeneities of various fields
are washed away, thereby providing a microscopic explanation of the existence of cosmological (time-dependent, homogeneous to leading order) backgrounds, which have been used in \cite{bms1,bms2} to discuss the RVM-like dynamical inflation. On the other hand, the second RVM-inflationary phase, implies that
any remnant of massive gravitinos or DW from the early phase of
the Universe, is washed out during the second inflationary period, at the
end of which only KR (or other stringy) axion backgrounds remain, while chiral matter is generated~\cite{bms1}.

For completeness, we mention at this point, that, as discussed in
\cite{sugraRVM}, this first inflation can also be described within the RVM unifying effective low-energy framework, which in this way can connect the Big-Bang to the present era of a string-inspired Cosmology. This RVM description also explains the absence of an initial singularity, which at a microscopic level is attributed to higher curvature terms in the string effective action~\cite{art}. \\

\section{The emergence of RVM  and the scale of Inflation \label{sec:inflscale}}

As mentioned previously, stiff axion matter becomes subdominant once the
GW condense in the gravitational anomalous terms in the action. Domain walls, as discussed above, and GW co exist with the stiff axion matter,
but once the condensation of GW becomes a dominant phase, then the situation drastically changes.
The condensation of GW { involves only the ground-state KR axion configuration \eqref{bfield}.} {As we shall show below in
subsection \ref{sec:eossrvm}, this phase}, {which succeeds the stiff-axion-dominance phase (cf. Fig.~\ref{fig:Hevolrvm})} {is characterised
by an equation of state of an RVM type  \eqref{rvmeos} ({i.e. of vacuum type with running vacuum}), due to the dominance of
the non-linear physics effects of the GW condensate.}

 {The GW-induced condensate of the anomalous interactions of the KR
axion with the gravitational Chern-Simons terms  is denoted by $ \langle \overline b \, R_{\mu\mu\rho\sigma}\, \widetilde R^{\mu\nu\rho\sigma} \rangle $. During inflation,  for which $H \simeq H_I = {\rm constant}$, it contributes  an effective cosmological-constant-type term in the effective action which reads as follows\cite{bms1,bms2}:}
\begin{align}\label{lambda}
\mathcal S_\Lambda  &=
\sqrt{\frac{2}{3}}\,
\frac{\alpha^\prime}{96 \, \kappa} \, \int d^4 x \sqrt{-g} \, \langle \overline b \, R_{\mu\mu\rho\sigma}\, \widetilde R^{\mu\nu\rho\sigma} \rangle  \equiv  -  \int d^4x \, \sqrt{-g} \, \frac{\Lambda (H)}{\kappa^2} \nonumber \\ & \simeq   \int d^4 x \, \sqrt{-g}\, \Big(5.86 \times 10^{-5}  \, \Big(\frac{\mu}{M_s}\Big)^4 \, \sqrt{2\, \epsilon} \,
\Big[\frac{\overline b(0)}{\MPl} + \sqrt{2\, \epsilon} \,  \mathcal N_e \Big] \, H^4 \Big)\,,
\end{align}
{where  $\epsilon$ is  the slow-roll parameter of inflation}, $\mu$
is an Ultra-Violet cutoff of the graviton modes, entering the
computation of the GW condensate~\cite{bms1,stephon}, and the notation $\simeq
$ indicates an order of magnitude estimate.
 {Notice that}  we have made use of the fact that $H\, t $ is bounded from above by $(H\, t)_{\rm max}$, the latter evaluated at the end of the inflationary period, that is
\begin{align}\label{efold}
(H\, t)_{\rm max} = H\, t_{\rm end} \sim {\mathcal N}_e = 60-70~,
\end{align}
with ${\mathcal N}_e$ the number of e-foldings. We stress again that {the notation $\Lambda (H)$ in Eq.,\eqref{lambda} implies that the effective cosmological term} is (approximately) constant during the de Sitter phase, in which the Hubble parameter is approximately constant, $H \simeq H_I$. { It may be appropriate to remind once more the reader that in Starobinsky inflation~\cite{staro} it is not $H$ but its cosmic time derivative, $\dot{H}$, which remains approximately constant during the inflationary period --  see \cite{ms21} for a brief comparative study.}  {In  \cite{bms1,bms2} we have set the slow-roll parameter in the
ballpark of}
\begin{align}\label{eps}
\epsilon \sim 10^{-2}~,
\end{align}
in agreement with conventional single-field inflationary phenomenology~\cite{bms1} { and consistent with the Planck data\,\cite{Planck}}. But in our RVM dynamical inflation, which is characterised by the absence of a conventional inflaton, much smaller $\epsilon$ are induced, as we shall discuss below. As long as $0 < \epsilon < 1$, successful inflation is guaranteed. So for the moment we keep $0 < \epsilon < 1$ as a parameter to be determined.

{We next remark that, as discussed in \cite{bms1,bms2}, to ensure constancy ({and not dilution}) of the gravitational anomaly condensate during inflation, and at the same time secure constancy of the resulting axion background after inflation, i.e. the fulfilment of  Eq.\,\eqref{eq:bfluxconst}, one must arrange that the following condition holds:}
\begin{align}\label{muH}
\frac{\mu}{M_s} \simeq 15 \, \Big(\frac{M_{\rm Pl}}{H}\Big)^{1/2}\,.
\end{align}
This relation should not be viewed as implying that the cutoff $\mu$ varies as a function of $H$, it is rather an {\it order of magnitude} constraint for $\mu$, for fixed $H$, which guarantees the {nonvanishing} constancy of the gravitational anomaly condensate during inflation, for which $H
\simeq H_I$.

\subsection{Transplanckian censorship hypothesis }\label{sec:transplank}

{The above relation can now be combined with a fashionable conjecture in the context strings from where to derive an interesting correlation
between the  string scale and the inflationary scale. In fact,  according
to the  Transplanckian Censorship Hypothesis (TCH) or conjecture~\cite{TCH}, which was  recently explored in connection to the swampland program of string theories~\cite{swampland}, no effective field theory
emerging from superstring theory can lead to a regime where fluctuation modes which were initially trans-Planckian ever exit the Hubble radius. If
we adopt the TCH for $\mu$ we can assert  that in an effective low energy
field theory we must actually have $\mu \simeq M_{\rm Pl}$, so that the effect of the transplanckian graviton modes are not considered in our low-energy approach. It is remarkable that in the context of such conjecture
it is possible  to determine the scale of inflation from \eqref{muH} as a
function of the string scale $M_s$. We find}
\be\label{inflscale}
H_I \simeq \frac{225 \, M_s^2}{\MPl}~.
\ee
{For example, for  $M_s\sim 10^{-2}\MPl$ this gives $H_I\sim 10^{-2}\MPl$. However, as we shall see $H_I$ is actually much lower than that and hence $M_s$ is reduced too.}
We next remark that, in our conventions~\cite{bms1}, \eqref{lambda} yields a positive contribution to the vacuum energy, as required for consistency of the model, on account of the Friedman equation, for sufficiently negative
$\overline b(0) < 0$, such that  $\frac{\overline b(0)}{\MPl} + \sqrt{2\,
\epsilon} \,  \mathcal N_e  < 0$.
In that case,
to ensure {\it approximate constancy} of the term inside the parentheses in the right-hand side of \eqref{lambda}, we imposed in
\cite{bms1,bms2} the condition
\begin{align}\label{b0bound}
\frac{|\overline b(0)|}{\MPl} \, {\gg}  \, \sqrt{2\, \epsilon} \,  \mathcal N_e = \mathcal O(10), \quad \overline b(0) < 0,
\end{align}
taking into account \eqref{efold}, \eqref{eps}.
{Actually, {the necessity of the above condition} \eqref{b0bound} becomes
apparent in view of the negativity of $\overline b(0) < 0$, and
Eq.~ \eqref{bfield}, so as to ensure that
during the entire inflationary era the KR field remains approximately constant in order of magnitude.
If, for instance, one had simply  $\frac{|\overline b(0)|}{\MPl} \, {\simeq}  \, \sqrt{2\, \epsilon} \,\mathcal N_e $, then the KR field, and consequently the
condensate \eqref{lambda}  would vary from a large negative value at the onset of inflation, to almost zero value at its end. What \eqref{b0bound}
implies is that one can only have configurations such that $-\frac{|\overline b(0)|}{\MPl} \, +  \, \sqrt{2\, \epsilon} \, \mathcal N_e
= - \tilde \xi \, \frac{|\overline b(0)|}{\MPl}$, with $\tilde \xi > 0$ a fraction of order 1, so that, in order of magnitude, the condensate is independent of time
during the RVM inflation.}

\subsection{Gravitational-Anomaly condensates}\label{eq:Condensates}

{ Because in the RVM the Hubble parameter remains essentially constant during inflation,  $H \simeq H_I$, the total energy density of the stringy-RVM vacuum taking into account the contributions
from the gravitational anomalous Chern-Simons terms and the condensate term \eqref{lambda} in the effective action, can be numerically estimated during the inflationary period and can be conveniently expressed as follows:
\begin{align}\label{toten}
0 <  \rho_{\rm total} &  \simeq
3\kappa^{-4} \, \Big[ \nu\, \Big(\kappa\, H \Big)^2
+ \alpha \, \left(\kappa\, H \right)^4 \Big]  = 3\kappa^{-4} \, \Big[ c_1 \,
+ c_2 \, \Big] \, \Big(\frac{\mu}{M_s}\Big)^4 \, \left(\kappa\, H_I \right)^4 \nonumber \\
&\simeq 3\kappa^{-4} \, \Big[ c_1 \,
+ c_2 \, \Big] \, (15)^4 \, \left(\kappa\, H_I \right)^2\,,
\end{align}
{where we have defined two numerical coefficients both depending of
the slow-roll parameter:}
\be\label{eq:c1c2}
c_1  = -{0.34}\,  \times 10^{-5}\, \epsilon , \,\,\ \
 c_2 = \frac{\sqrt{2}}{3} \, |\overline b(0)| \, \kappa \, \times {5.86\, \times} \, 10^{-5} \, \sqrt{\epsilon} \, \simeq 2.8 \times   \, 10^{-5} \, \sqrt{\epsilon} \,
|\overline b(0)| \, \kappa ~.
\ee
{ The second parameter also depends on the initial value of the mean KR field during inflation, $\overline b(0)$, and we will elaborate further on
it. Let us  mention that apart from the almost constancy of the Hubble parameter during inflation,  we have used \eqref{muH}, and, at this stage, assumed that $0 < \epsilon < 1$ (actually, $\epsilon\ll1$, see later on, \eqref{eps2}, for a justification of this assumption).
As indicated, the above expression is a gross estimate of the total vacuum energy density  during the inflationary stage, but as  it is obvious from the general structure or the RVM, Eq.\,\eqref{rLRVM}, the RVM vacuum density is a dynamical quantity which is dominated  by the two contributions $\sim H^2$ and $\sim  H^4$ at high energy. The details justifying why the balance between these two dynamical terms lead to the above numerical
estimate during the inflationary time in our stringy-RVM framework  have been put forward in  ~\cite{bms1,bms2}.} {In subsection \eqref{sec:eossrvm} below, we shall also compute the vacuum equation of state for this phase, demonstrating explicitly that it is that of the RVM vacuum \eqref{rvmeos}.}

The form \eqref{toten} is of RVM type, with the important difference, however, that the coefficient of the $H^2$ term is {\it negative}, as a consequence of the negative contributions of the gravitational anomalies. This is to be contrasted with the conventional RVM where this coefficient is
positive ({such a situation is however met also in the post-RVM inflationary era of the
stringy RVM model~\cite{bms1}).  The reader should also notice
that, in order of magnitude, the use of the condition \eqref{muH} in \eqref{toten} implies that both terms in $\rho_{\rm total} $ are proportional
to $H_I^2$ during inflation. This is not the case in the general RVM.  Nonetheless, the $H^4$ term in the RVM dominates, and the total energy turns out positive, as required for consistency with the Friedman equation. In fact, from the above formula we find
\be\label{eq:ratioc1c2}
\frac{|c_1|}{c_2}\simeq \frac{\sqrt{\epsilon}}{{10} \, |b(0)|\, \kappa}\ll {\frac{\sqrt{\epsilon}}{100 }} \ll 1\,,
\ee
where we used Eq.\,\eqref{b0bound}. As we shall see, $|\epsilon|\ll1$, a fact already known from the slow roll condition\cite{bms2,bms3}, but  which will get further qualified below (\eqref{eps2}).}

{The following point is now in order.  On account that the constant KR background  $|\overline b(0)|$ is allowed to take on transplanckian values ({\it cf.} \eqref{b0bound}), the alert reader might worry  if this is compatible with the validity of the effective low-energy field theory below Planck scale.  Interestingly enough, the answer is positive.  First,  during the inflationary phase our effective theory depends only on derivatives of the massless KR axion, $\dot b$, which in the light  of \eqref{dotearly}, is sufficiently smaller than $\MPl^2$ to justify the validity of the effective field theory \eqref{sea4}. This comes about because
\be\label{eq:dotbsmall}
\frac{\dot{b}}{\MPl^2}=\sqrt{2\epsilon}\ \frac{H}{\MPl}<\sqrt{2\epsilon}\  10^{-4}\ll1\,,
\ee
where we have used  $0<\epsilon<1$ (in fact, $0<\epsilon\ll1$, as we shall see later on, \eqref{eps2}) and the fact that  during inflation $H/\MPl\lesssim 10^{-4}$.
Second, despite of the fact that  the anomalous condensate value in  Eq.\,\eqref{toten} does depend on $b(0)$ itself through the value of the coefficient $c_2$ above,  this does not violate the TCH either since the condensate term, being proportional to $H^4$, also assumes subplanckian values owing to  $H\lesssim H_I<M_P$ (see Sec.\,\ref{Sec:ScaleInflation} for details);  and this holds good notwithstanding the Planck size fluctuations of the KR field. Therefore, in all cases we meet  compatibility with the effective theory and with current phenomenology.   This concludes our proof  that  all the terms in the effective action of our framework assume
sub-planckian values, thereby perfectly respecting the TCH.}

 {The result \eqref{eq:ratioc1c2} is important for yielding RVM-de Sitter
dominance after the GW condensation.
{In the next section we show} that the equation of state characterising the energy and pressure density of the vacuum, denoted by
$\rho_{\rm  total}$ and $p_{\rm total}$, respectively, is that of
 the RVM  \eqref{rvmeos}.
 First we remind the reader that in such vacuum contributions, the KR axion field is given by \eqref{bfield}, as it finds itself in its ground state, that is, it satisfies its classical equation of motion.}

 \subsection{Equation of state  of the  stiff medium with anomalies and GW-condensate \label{sec:eossrvm}}

{To calculate $\rho_{\rm total}$, $p_{\rm total}$
we use the expression of the total stress energy tensor~\cite{bms1}, which includes the interactions of the ground state KR field with the anomalies. The gCS terms are associated with
the Cotton tensor ${\mathcal C}_{\mu\nu}$ in the way explained in \cite{bms1,bms2}.
Hence
\begin{align}\label{stiffvacuumexcit}
\rho_{\rm total} =  \rho^b + \rho^{\rm gCS} + \rho^{\rm condensate} , \quad
p_{\rm total} = p^b + p^{\rm gCS} + p^{\rm condensate},
\end{align}
where we split the ``vacuum'' contributions into parts associated with the ground state of the KR axion field $b$,
satisfying a stiff equation of state
\be\label{stiffgr}
p^b = + \rho^b,
\ee
the Cotton tensor, for which we use the abbreviation ``gCS'', as per the notation of \cite{bms1},
and the condensate \eqref{lambda}, satisfying a de Sitter equation of state:
\be\label{dS}
p^{\rm condensate} = - \rho^{\rm condensate}.
\ee}

{We next note that the gravitational tracelessness of the Cotton tensor \eqref{trace}
implies a radiation-like relation for the corresponding contributions to energy density and pressure,
\begin{align}\label{rad}
p^{\rm gCS} = \frac{1}{3} \, \rho^{\rm gCS}~,
\end{align}
where the pressure density terms $p^{\rm gCS} $ are associated with the spatial diagonal components of the Cotton tensor \eqref{cotton},
$C_{ii}$, no sum oiver $i$, whilst the energy-density $\rho^{\rm gCS}$ terms are linked to its temporal components $C_{00}$~\cite{bms1}.}

{In calculating \eqref{trace}, which leads to \eqref{rad},
one takes into account the GW perturbations, which condense in the de Sitter phase, to yield
a non zero Cotton tensor \eqref{cotton}~\cite{bms1}. However, we are working with weak perturbations, as naturally expected from the fact that we are dealing with quantum gravity effects, which implies that, to leading order, $g^{\mu\nu}$ in \eqref{rad} will be taken to be the FLRW background metric, diagonal in its elements. This leads to \eqref{rad}. }

{As shown
in \cite{bms1}, $\rho_{\rm Cotton~tensor} < 0$, hence \eqref{rad} would also imply negative `pressure' contributions by the
Cotton anomalous terms $p_{\rm Cotton~tensor} < 0$.
The calculation of \cite{bms1} has yielded the relation
\begin{align}\label{csb}
\rho^b=-\frac{2}{3} \, \rho^{\rm gCS} > 0,
\end{align}
which implies for the sum of the vacuum contribution of the $b$ field plus those of the gravitational Chern-Simons terms would {satisfy}:
\begin{align}\label{bgCS}
\rho^b + \rho^{\rm gCS} &= \frac{1}{3}\, \rho^{\rm gCS} < 0~, \nonumber
\\
p^b + p^{\rm gCS} &= \rho^b + \frac{1}{3} \, \rho^{\rm gCS} = -\frac{1}{3}\, \rho^{\rm gCS} = - (\rho^b + \rho^{\rm gCS})>0~,
\end{align}
and thus they,  independently of $\rho^{\rm condensate}$, satisfy the RVM
equation of state \eqref{rvmeos}. But the total energy of these two contributions is negative} { and hence they do not correspond to standard vacuum since the total pressure is positive. Such kind of exotic state is, however,  only transitory until the true (dynamical) vacuum state represented by the RVM is attained. In the next section we comment on the
various forms of EoS in cosmology, some of them being rather exotic, but
not impossible to meet in different stages of the universe's evolution. }

\subsection{Vacuum versus phantom-vacuum and phantom matter }

{Some taxonomy may be useful at this point to categorize the different forms of equation of state and to find a place for  the exotic EoS
species \eqref{bgCS} }. {We have seen that the anomaly terms contribute negatively to the energy density of
the coupled system $b$-field-gravitational-Chern-Simons-terms ({\it cf.} \eqref{sea4}), and dominate over the positive energy density of the free KR-axion $b$-field terms.
This fact makes this situation reminiscent of the so-called {\it phantom ``matter''} (PM) system, introduced in \cite{phantmat1}, and elaborated further in \cite{phantmat2}. Indeed, PM in those works is defined as a substance  $X$ (not related to matter, but rather a component of dark energy) satisfying $p > -\rho$,  but with $\rho < 0$ (see fig.~\ref{fig:PM})\footnote{Phantom matter is not a completely unusual theoretical concept; it resembles the kind of exotic matter  with negative energy  (thus violating the weak energy condition) that one needs in order to stabilize certain kinds of `traversable wormholes'\,\cite{Thorn1988}.},
where the blue sector marked "P" is the usual phantom energy (characterised by $p=w \, \rho$, $\rho > 0$, {$w < -1$. The limiting case $w=-1$ is out of the phantom region,  as it corresponding to the vacuum
energy EoS,  whether rigid or dynamical, it includes the cosmological constant term, the  standard de Sitter cases and also the RVM as dynamical vacuum}), whereas the PM region lies diametrically opposite to it on the top left
part).
The PM region satisfies $p=w\rho$  with
$w \le -1 $ and $\rho < 0$, hence $p > 0$.  PM was called phantom ``matter'' in \cite{phantmat1}, because, in contrast to the usual phantom energy, PM does satisfy the strong energy conditions, $p + \rho \ge 0, p + 3p \ge 0$ similar to ordinary matter.
In \cite{phantmat1,phantmat2} the dynamics of the PM substance $X$ has been studied in the context of the RVM of Cosmology, with the running vacuum energy exchanging energy with X. In the approach of \cite{phantmat1,phantmat2}, because the PM substance was taken to be independent of the RVM,
and the PM is {\it attractive}, like ordinary matter, in contrast to phantom energies, which lead to effectively repulsive forces in the Universe,
one could also bring the current expansion of the Universe to a halt at some point in the future.  {This feature may be helpful as a possible explanation of the cosmic coincidence problem, as proposed in\,\cite{phantmat1}}. }

\begin{figure}[t]
\begin{center}
\includegraphics[width=0.8\textwidth]{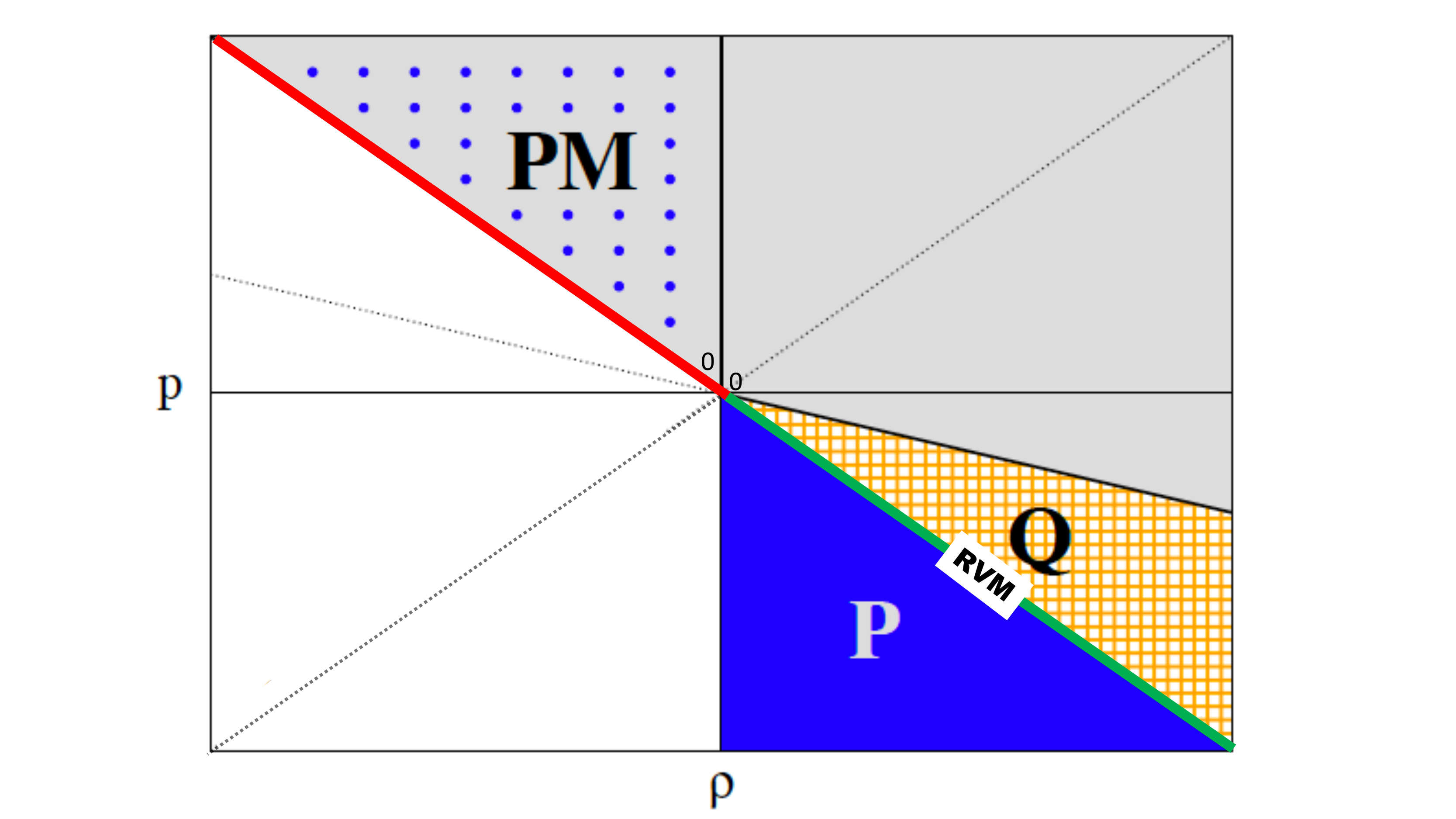}
\end{center}
\caption{{\it A diagram of pressure density versus energy density corresponding to various fluids with equation of state $w$ {of interest
in cosmology}, including standard quintessence (Q: { $-1 < w < -1/3$}), phantom energy (P:  {$w <-1$, $\rho > 0$}),
phantom ``matter'' {(PM:  $w < -1, \rho < 0$),  rigid or dynamical (RVM) vacuum ($w=-1, \, \rho>0$:  green line in the antidiagonal) and phantom
vacuum ($w= -1,\, \rho < 0$: red line in the antidiagonal)}.
The gray-shaded regions (with or without dots) correspond to regions for which the strong energy conditions are satisfied, $\rho + p \ge 0, \rho +
3p \ge 0$.
The string-inspired gravitational field theory with anomalies we are studying in this work  lies exclusively on the diagonal, $w=-1$, and, depending on {the value and sign of the KR axion field at the onset of  inflation, $\overline b(0)$, it can pass from the red line (phantom vacuum)  to
the blue line, {\it i.e.}  dynamical true vacuum (RVM-inflation), and vice versa.  Adapted from ref.~\cite{phantmat1}}.}}
\label{fig:PM}
\end{figure}

{Having briefly summarized the taxonomy of possible  EoS in cosmology,  we can see that  the  case \eqref{bgCS}  that we have encountered
in the context of the string-inspired gravitational theories with
anomalies coupled to KR axion fields, \eqref{sea4}, corresponds to the limiting situation of PM for which  $p=-\rho>0$ (exactly),  i.e. $w=-1$
 with $\rho <0$ (represented by the red line in our Fig. 2), whereas general PM admits  $w<-1$ with positive pressure.  The exact limiting case $p=-\rho>0$  can be called `phantom vacuum' for obvious reasons.
Notwithstanding  its sharing some features similar to the PM (in particular the fulfillment of the strong energy condition),  the phantom vacuum case that we have found has also important physical differences, which we now come
to discuss}. {In our model, the system that could play the r\^ole of the PM
consists of the KR axion in its ground state configuration \eqref{bfield}, coupled to the gravitational anomaly terms. In the presence of GW perturbations that condense, the latter lead  to a non-zero Cotton tensor \eqref{cotton}, which contributes negatively to the energy density of the coupled system, dominating the positive contributions coming from the kinetic terms of the KR  axion background configurations \eqref{bfield}. However these are not the only terms generated by the condensation of GW. There
is also the cosmological constant term \eqref{lambda}, which is of an approximately de Sitter type, provided the condition \eqref{b0bound} is satisfied. In this way, the de Sitter contribution
dominates, and leads to an overall positive $\rho_{\rm total}$ \eqref{toten}, since the Cotton-tensor negative contributions to the energy density, become now subdominant. One is tempted to identify the Cotton-tensor-KR-axion coupled system with {a state of phantom vacuum}, and the de-Sitter
type due to the condensate term \eqref{lambda} with the RVM ({that carries true, but dynamical, vacuum}), which then dominates {over the phantom vacuum transitory period}, leading finally to the true vacuum, as in the model of \cite{phantmat1,phantmat2}. However, this identification is not exactly accurate, given that all three of the above contributions occur simultaneously in the model, once GW condense, and thus they cannot be considered as pertaining to different systems, as
it was the case of the X substance in the PM model of \cite{phantmat1,phantmat2}. {Nonetheless,} {the comparison is pertinent in that
all of three components (two combining into phantom-vacuum and the third one being true vacuum)  finally conjoin so as to  constitute an overall  dynamical vacuum configuration, of RVM type~\cite{bms1}, in the sense of the total energy density and pressure satisfying \eqref{rvmeos}, with the
total energy
density of the fluid being positive \eqref{toten}}. This is the true meaning of \eqref{bgCS}, which constitutes only part of the whole system, the
latter involving also the condensate \eqref{lambda}, which thus cannot be
considered separately.}

{To summarize, despite} {the fact that} {the zoo of EoS is rather rich in
cosmology and the most exotic forms are all possible in the very early universe, we have seen that our stringy system naturally tends to form  a  true vacuum state with positive energy.  It starts with vacuum (at the hill-top inflationary era of Fig. \ref{fig:Hevolrvm},
see Sec. \ref{sec:modern} for details), and moves to a phantom-vacuum regime made of stiff matter and anomalies,  mixed with a dominant GW-condensate \eqref{lambda} resulting in  the RVM-vacuum with EoS \eqref{rvmeos}. This is clear from the fact that $c_2\gg c_1$  in \eqref{eq:ratioc1c2}, which
implies that $\rho^{\rm total}$ is dominated by positive (running) vacuum contributions giving rise to the emergence of  RVM inflation~\cite{bms1}. This is the complete and consistent picture, which we are  unveiling  for the first time in the present work.}

\subsection{The scale of the RVM inflation and the string scale}\label{Sec:ScaleInflation}

{The initial value and sign of the KR field at the onset of the RVM inflation, $\overline b(0)$, is crucial to determine whether we lie in the phantom vacuum region or in the true vacuum one (RVM).} { If it
was positive rather than negative, then the energy density of the term $\eqref{lambda}$  {would also be positive, thus leading to a negative energy density situation of phantom-vacuum type, which would entail $\Lambda(H)<0$, the system being characterised by a limiting case of PM, with $\rho_{\rm total} <0$ and $p_{\rm total}=-\rho_{\rm total}$} (this latter relation is due to \eqref{bgCS}, combined with the RVM equation of
state \eqref{rvmeos} for the analogue of the condensate \eqref{lambda} with $\overline b(0) >0$).
Hence, depending on the range of values of the parameter $\overline b(0)$
one may pass from a {phantom vacuum situation with $w=-1$, to a true vacuum one RVM}  (passing from the red to green lines in fig.~\ref{fig:PM}). It is in this latter sense that the considerations of \cite{phantmat1} find an application in our string-inspired cosmology.}

{After this important digression, we now come back to our analysis.}
If one assumes that the Friedman equation is satisfied when the condensate  \eqref{toten} is formed\footnote{The assumption that Einstein's classical equations of motion are satisfied by the condensate \eqref{lambda} might not be true
given that we deal with an unknown path integral of full stringy quantum gravity, involving all higher-order curvature and covariant derivative corrections. Here, we explore the physical consequences of such an assumption, and as we show, these are not unreasonable.}, then one would obtain (for ~$0 < \epsilon \ll 1$, which, as we shall see, is a self consistent assumption):
\begin{align}\label{fried}
H_I^2\simeq {\frac{\kappa^2}{3}\rho_{\rm total}\simeq \kappa^{-2} c_2 (15)^4 \, \left(\kappa\, H_I \right)^2}\simeq 1.4 \, \sqrt{\epsilon}\,
\frac{|\overline b(0)|}{M_{\rm Pl}} \, H_I^2
 \quad \Rightarrow \quad  \frac{|\overline b(0)|}{M_{\rm Pl}}  \simeq 0.71 \, \epsilon^{-1/2}
\end{align}
{where we have neglected $|c_1|\ll c_2$  thanks to the relation \eqref{eq:ratioc1c2} and used the numerical value of $c_2$ from  \eqref{toten}.  On account of the bound \eqref{b0bound},  it follows that}
\be\label{eps2}
\epsilon \lesssim  0.5 \, \mathcal N^{-1}\sim 7\times 10^{-3}\,,
\ee
{where $\mathcal{N} = 60-70$.  Thus we confirm that $|\epsilon|\ll1$, as mentioned above.  On the other hand, if we compare equations \eqref{rLRVM} and  \eqref{toten}-\eqref{eq:c1c2} we find that the value of $\nu$ during the primeval inflationary stage {would be $\nu\simeq c_1 (15)^4\simeq -1.2\times 10^{-3}$}, which is in the expected range but
with opposite sign as compared to the low energy value obtained from  fitting the cosmological data\,\cite{rvmpheno1, rvmpheno2,rvmpheno3}.  This is because of the anomaly influence,  which  is  well under way in the inflationary epoch.  Once we exit the de Sitter phase and enter the normal radiation phase
of the usual FLRW regime, the gravitational anomaly ceases to act (it vanishes identically for FLRW spacetime\cite{bms1,bms2}) .  Then,  the low energy effective value of the  $H^2$-coefficient    in our stringy-RVM approach  is to be associated with the contributions of the $b$-axion field stress tensor only, with no counter effect from the Chern-Simons term (causing the gravitational anomaly). From the definition of slow-roll parameter in scalar field driven inflation, Eq.\,\eqref{eq:SlowRoll}, and observing that the kinetic term of the KR field ($\frac12\,\dot{b}^2=\epsilon H^2/\kappa^2$) contributes to the $\nu$-term of \eqref{rLRVM}, we find
\begin{equation}\label{nub}
\nu_b= \frac{\epsilon}{3}\simeq 3 \times 10^{-3} > 0\,,
\end{equation}
{where we have used \eqref{eps2} and called this (post-inflationary) contribution $\nu_b$,  as it is only associated to the KR field $b$.
As it turns out}, the post-inflationary value of $\nu$  is of the same order of magnitude as the full $\nu$ value (previously determined)  when the anomaly was also present,  but positive.  This fact obviously evinces the great influence of the anomaly part during inflation. {But at
the same time  because that part ceases to act during the FLRW era, we find that  the order of magnitude and positive sign of $\nu$ obtained in the usual FLRW regime, i.e. Eq.\,\eqref{nub},  is in perfect agreement with
the fitting  results obtained for $\nu$ using the cosmological data, which  confirms  the excellent phenomenological performance of the
RVM} as a dynamical vacuum framework able to improve the description of the cosmological data\,\cite{rvmpheno1, rvmpheno2,rvmpheno3} and  also capable of
alleviating the $H_0$ and $\sigma_8$ tensions\,\cite{tensions} afflicting
the $\CC$CDM, as recently shown in\,\cite{EPLtensions}. See also \cite{Valentino1,Valentino2}, and references therein, for a summary of the many existing models trying to cope with these tensions. }

{We have shown in Sec.\,\ref{eq:Condensates} that  all the terms
in the effective action of our framework assume subplanckian values, whereby  respecting the TCH.
However, the inflationary scale $H_I$ must take on subplanckian values as
well. Let us check it explicitly.}

We first note that, from a theoretical point of view, {there is no lower bound on $H_I$}, given that $M_{\rm Pl}/|\overline b(0)|$,
{and hence $\epsilon$ ({\it cf.} \eqref{fried}),
could be arbitrarily small}.  {As we have seen, a transplanckian
$|\overline b(0)|$ is compatible with the TCH, given that
the effective gravitational action that contains the string-model independent KR axion $b(t)$, involves only its derivatives $\partial_\mu b$ which assume subplanckian values, {and the GW condensate is also subplankian}.} From a phenomenological point of view, however, one might think of imposing
such a lower bound  by the requirement for realising {successful BBN} at temperatures of order $T \sim 1$~MeV~\cite{bbn}.
This happens in conventional models  with single-field inflation, which involve reheating of the Universe during the exit from inflation at a given temperature $T_R$, which thus has to be less or of order the inflation scale. In \cite{bbn} models with low reheating temperatures are considered, which lead to the aforementioned lower bound on $H_I$. In our RVM, {there is no conventional reheating of the Universe \,\cite{RVMinfl,JSPRev2015,GRF2015,solaentropy}}, hence the considerations of \cite{bbn}
need to be rethought.

In what follows, we use phenomenological data to impose a restriction on the size of $|\overline b(0)|$, which is treated as a
phenomenological parameter. {The value of the inflationary scale
$H_I$ can be estimated from the anisotropies of the CMB\,\cite{JSPRev2015}.
According to the CMB observations, the (primordial) spectrum ${\cal
P}_{\zeta}$ of the curvature perturbation  at
typical ``pivot'' length scales $k_0^{-1}$ (of a few tens to
hundreds of Mpc) is ${\cal P}_{\zeta}^{1/2}\simeq 5\times 10^{-5}$. Using
{\it single-field slow-roll models} of inflation, the scale of inflation $H_I$  can then be expressed in terms of the ratio $r$ of tensor to scalar perturbations as follows\,\cite{JSPRev2015}:}
 \begin{align}\label{rstar}
 H_I = 8 \times 10^{13} \, \sqrt{\frac{r}{0.1}} \, {\rm GeV} = 3.28 \times 10^{-5} \, \sqrt{\frac{r}{0.1}} \, M_{\rm Pl}~.
\end{align}
From the joint analysis of of BICEP2/Keck Array and Planck data~\cite{bicep2,Planck} the current experimental upper bound of $r$, $r^{\rm max}$,
is
\begin{align}\label{rupper}
r \, < \, r^{\rm max} = 0.12~.
\end{align}
{Notice that if there is a  Grand Unified Theory (GUT) scale, $M_X$, the latter is related to $H_I$ roughly  through $H_I^2=(\kappa^2/3) M_X^4$,
which indeed implies that  $M_X$ is in the  expected order of magnitude: $M_X\simeq 1.8\times 10^{16}$ GeV for  $r$  near the upper  bound  $ r^{\rm max}\simeq 0.1.$}

We stress once more that no strict lower bound exists except the requirement for realising {successful BBN} at temperature of order $T \sim 1$~MeV~\cite{bbn}, which however, as already mentioned, cannot be applied straightforwardly in our RVM case. Hence, there is a huge gap between the scales at which the dynamics of the
Universe is understood and inflation.\footnote{An interesting theoretical
suggestion of restricting the inflationary scale by means of dark matter abundances  in models  where the dark matter is a self-interacting weakly
interacting scalar is suggested in \cite{enqvist}. In our model~\cite{bms1,bms2}, during inflation, the only scalar fields available are the KR axion, which has no potential during that era,  and the condensate degree of freedom, whose potential is not known, although we expected the condensate to have gravitational-in-origin self interactions. The KR axion can acquire self interactions only after inflation, e.g. during the QCD era and become a dark-matter candidate. Moreover, we may have more than one type of stringy axions, that interact with the KR axion. Hence the scenario of \cite{enqvist} for restricting the scale of RTVM-inflation cannot be straightforwardly realised in our model.}

From \eqref{inflscale}, \eqref{rstar} and \eqref{rupper} we obtain:
\begin{align}\label{posit2}
\frac{H^{\rm max}_I}{M_{\rm Pl}} \simeq  \, 3.28 \times 10^{-5} \, \sqrt{\frac{r^{\rm max}}{0.1}}  \simeq 3.59 \times 10^{-5} > \,
\frac{H_I}{M_{\rm Pl}}  \simeq 225 \frac{M_s^2}{M^2_{\rm Pl}}~,
\end{align}
implying
\be\label{msmpl}
M_s <  1.3 \times 10^{-4} \, M_{\rm Pl}\sim 10^{14}\, \text{GeV}\,.
\ee
{In the present  TCH context we would thus predict a string scale below the typical GUT scale $M_X\sim 10^{16}$ GeV.\footnote{{We recall at this stage that, according to the discussion in section \ref{sec:sRVM}, there is a mild dependence on this bound on the microscopic model, namely a dependence on the quantity $\xi$ entering the Green-Schwarz gravitational Chern-Simons counterterms \eqref{csterms}, which is proportional to the number $\mathcal N$ of microscopic chiral fermionic stringy d.o.f. circulating in the anomalous quantum loop diagrams. For the maximum expected $\mathcal N \sim 10^3$ in realistic string models~\cite{stephon}, one has $\xi \sim 6$, which increases, as explained in section \ref{sec:sRVM}, the bound \eqref{msmpl} by almost an order of magnitude.}}
Let us however recall that the string scale can be as low as a few TeV,  in principle, in models
with large extra dimensions.}
In this respect, let us stress that it is not possible in our approach to
determine theoretically the scale of the stringy RVM inflation without imposing the transplanckian censorship on the cutoff $\mu$, which led to \eqref{inflscale}. This is because the apparently
$H^4$ dependence of the second term in the right-hand-side of the expression \eqref{toten}, which is valid for general $H$, becomes equivalent to a $H_I^2$ during inflation  -- {with, however,  an enhanced coefficient $c_2$ as compared to $c_1$,  of course, see \eqref{eq:ratioc1c2}}
--  for which \eqref{muH} applies.
This result, is consistent with \eqref{b0bound}, and strengthens even more the assumption on the de Sitter nature of the condensate \eqref{lambda}
during the inflationary era, given that we may assume sufficiently transplanckian  values of $|\overline b(0)|$ such that
\eqref{lambda} remains practically unchanged during the entire duration of inflation. Quantum fluctuations of the condensate are then responsible for deviations from scale invariance, providing a novel mechanism for cosmological perturbations.

\section{Discussion: Modern Era Phenomenology of the Stringy RVM \label{sec:modern}}

{In this section we make some important remarks concerning the phenomenology of our stringy RVM in the current cosmological era.
We remark that, in the modern era, the phenomenology of the stringy RVM is similar to the conventional RVM~\cite{rvmpheno1,rvmpheno2,rvmpheno3,EPLtensions}, as far as a deviation from the predictions of the $\Lambda$CDM
model is concerned.
We commence the discussion by first discussing the phenomenology of the conventional RVM, specifically the type II RVM.}

\subsection{Type II RVM as an ideal candidate to resolve the $H_0$ and $\sigma_8$ tensions}

{Due to the general form of the vacuum energy density \eqref{rLRVM}, for the small current values of $H$ it boils down to
\begin{equation}\label{eq:RVMcurrent}
  \rho_{\rm RVM} = \frac{3}{\kappa^2} (c_0 + \nu_0 H^2)\,.
\end{equation}
Obviously, since we find  $|\nu_0|\ll1$ (both theoretically and as a result of the global fits to the cosmological data~\cite{rvmpheno1,rvmpheno2,rvmpheno3,EPLtensions})}\footnote{{Here we use $\nu_0$ to distinguish it from other values we have assigned to this parameter in different cosmic epochs of the stringy version of the RVM.  The value $\nu_0$  applies to the current epoch and most likely in the entire post-inflationary period.
It should coincide in order of magnitude with our theoretical prediction
\eqref{nub}.  However, we cannot entirely exclude that other features of the post-inflationary era might have influenced also its ultimate value, which is in any case fitted to the data in the mentioned references. We shall not dwell further on this possibility here, though.} } {The behavior
of \eqref{eq:RVMcurrent} is that of a dominant cosmological constant term
$\Lambda_0=3c_0$ which is  corrected by a mild dynamical term $\nu_0 H^2$, with $\nu_0$ of order $10^{-3}$. This small dynamics, however, is sufficient to alleviate the tension associated to structure formation (viz. $\sigma_8$) and combined with a mild evolution of Newton's coupling (see below) it can address also the $H_0$ tension in a rather efficient way \cite{EPLtensions}.
Therefore, the  stringy RVM  benefits as well from the general ability of
the running vacuum model as a potential candidate for solving  the worrisome cosmological data tensions which persistently  afflict the $\CC$CDM~\cite{tensions,EPLtensions,Valentino1,Valentino2}.}

{All that said, we should emphasize that the stringy version of the RVM also makes specific predictions. It predicts  axionic dark matter, which is  provided
by the bulk flow of KR axions created at the end of the RVM inflationary period, Eq.\,\eqref{eq:bfluxconst}. These axions  may acquire non-perturbative masses} during the radiation or matter eras~\cite{bms2}. Moreover, the existence of Lorentz-
and CPT-violating backgrounds of the KR axion during
the radiation era, as a consequence of anomaly condensates,
which leads to leptogenesis~\cite{bms1,decesare} during the radiation era,
leave in principle testable traces to the CMB spectrum.

We hope we can come back to a detailed discussion of such topics  in a future work.
{Nonetheless, before closing we would like to make  some speculative remarks on the potential resolution of the data tensions $H_0$ and $\sigma_8$. }
{First of all, the structure \eqref{eq:RVMcurrent} can actually be more general.  Quantum corrections of the effective action of semiclassical gravity (classical gravity with quantum matter fields) indicate that the most general low-energy from is\,\cite{Cristian2020}
\begin{equation}\label{eq:RVMcurrentGeneral}
  \rho_{\rm RVM} = \frac{3}{\kappa^2} (c_0 + \nu_0 H^2 + \tilde{\nu}_0\dot{H})\,,
\end{equation}
where $\tilde{\nu}_0$ is another small (dimensionless) coefficient. The particular case $\tilde{\nu}_0=\nu_0/2$ was studied in \cite{EPLtensions} since then the linear combination of $H^2$ and $\dot{H}$ appearing in \eqref{eq:RVMcurrentGeneral} is proportional to the curvature scalar $R$. However, this is not relevant for the kind of considerations we want to make here.  What is important is that a generic low-energy RVM density of the form \eqref{eq:RVMcurrentGeneral} can be very useful to deal with the
tensions, in particular the $\sigma_8$ one because the dynamical increase
of $ \rho_{\rm RVM}$ in the past helps at the epoch of structure formation helps to suppress the formation of structure and is reflected in a lower value of $\sigma_8$ as compared to the $\CC$CDM~\cite{rvmpheno1,rvmpheno2,rvmpheno3,EPLtensions}.  But a new ingredient is still necessary in order to deal with the two tensions at a time.}

{That additional ingredient was highlighted in  \cite{EPLtensions}. In the latter work it has been shown that
the $H_0$ tension can be simultaneously alleviated with the $\sigma_8$ one
by allowing a mild (logarithmic) dependence on time of the gravitational constant, which in the notation of \cite{EPLtensions} is represented by the replacement of $\kappa^2$ by $\kappa^2/\varphi (t)$, with $t$ the cosmic time. The variable $\varphi$ mimics a Brans-Dicke field\,\cite{BD}, but it is not really so since it is not a field degree of freedom, just a convenient parameterization of the time evolution of Newton's $G$, specifically $G(t)\equiv G_N/\varphi$, where $G_N=1/(8\pi\MPl^2)$ is the local
Newton's constant\footnote{{The effect is similar to considering
a true BD model with a cosmological term\,~\cite{SolaBD}}.}.
Friedman equation then assumes the form:
\begin{align}\label{typeII}
H^2=\frac{\kappa^2}{3\varphi(t)} \Big[ \rho_m + \rho_{\rm RVM}\Big]=\frac{\kappa^2}{3\varphi(t)} \Big[ \rho_m +  C_0 + \frac{3}{\,\kappa^2} (\nu_0 H^2 + \tilde{\nu}_0\dot{H})\Big]\,,
\end{align}
with $C_0=3c_0/(8\pi G_N) $.  This constant would be the  traditional (rigid) CC term
$\Lambda$ if the vacuum would not be dynamical ($\nu_0=\tilde{\nu}_0=0$). The structure \eqref{typeII} is what it has been called type II RVM in \cite{EPLtensions} so as to distinguish it from the more conventional (type I RVM) in which the effective Newton's coupling is strictly constant.
The quantity $\rho_m$ denotes the total matter energy density, which is assumed conserved in the model of \cite{EPLtensions} (namely $\dot{\rho}_m
+ 3H\rho_m =0$, if restricting to the current epoch with pressureless matter). As noted, $\varphi(t)$ is {\it not} viewed as a fundamental field
(in contrast to the Brans-Dicke (BD) model~\cite{BD}), but merely incorporates an effective mild (logarithmic) time dependence of the gravitational constant. Indeed, the Bianchi identity for the curvature tensors implies
\begin{align}\label{typeIIBianchi}
\frac{\dot \varphi}{\varphi} = \frac{\dot \rho_{\rm RVM}}{\rho_m + \rho_{\rm RVM}\,}.
\end{align}
Solving for $\varphi$  with the help  \eqref{typeII} and \eqref{typeIIBianchi} one can show that its evolution is logarithmic  with the scale factor and that $\varphi<1$ at present\,\cite{EPLtensions} . As a result $H_0$  is larger than predicted by the $\CC$CDM, which is welcome for the $H_0$ tension.    So, as it turns out, the combined effect of dynamical vacuum energy from the RVM with the mild variation of $G$  is a possible clue
to fix to a large extend the two tensions $\sigma_8$ and $H_0$. }

{\subsection{Aiming at Type II RVM from  {supergravity}  and strings \label{sec:rvmIIstrings}}}

{A natural question arises at this point as to how one can incorporate
such a dependence in our stringy model, in order to address the issue of the Hubble tensions. We do not enter here in a detailed discussion on this, but we only give below some speculative remarks. A detailed analysis of this topic is reserved for a future publication.

\subsubsection{String-inspired Dilaton-dominance models}

An immediate thought to resemble \eqref{typeII} would be to incorporate a
time-dependent dilaton field at late eras, assuming dilaton dominance for
such epochs. This would make the situation resembling the BD model rather
than the type II RVM vacuum, which in any case from the point
of view of alleviation of tensions would be sufficient, in view of the above discussion. However, the important question, as far as string theory is concerned, concern the  microscopic string theory models that allow such a time-dependent (dimensionless) dilaton $\Phi(t)$ in the modern era, without affecting much the string phenomenology, given that the exponential of the dilaton is the string coupling $g_s = e^{\Phi(t)}$, and a time dependent string coupling
affects the phenomenological coupling constants of the low-energy field theory, which are proportional to $g_s$. Moreover, the precise time dependence of $\Phi(t)$ would depend on the dilaton potential, whose form is not known in general.}

{It is worthy of remarking that in some non-critical cosmological
string models~\cite{diamand}, where the dilaton is the so-called Liouville mode, one may consider (at tree-string-loop-level) exponential dilaton potentials of quintessence type, $V(\Phi) \sim V_0 \, e^{-{\rm c_\Phi}\, \Phi}$, where $V_0\, \kappa^{2}$, ${\rm c_\Phi}$ are dimensionless constants. In microscopic string theories, the form of $V$ is not known, as it
may be the result of non-perturbative string-quantum corrections, which at present are not known in closed form. In studies like the ones mentioned above, one adopts phenomenological forms, whose parameters can be fixed
by fits to the data~\cite{diamand,lahanas}. In fact one may even assumes different constants $c_\Phi$ at different late eras of the Universe, or put it differently, more complicated dilaton potentials.}.

{For the moment, we mention that, if one assumes dilaton dominance at
late eras, for such models, then the
cosmological Einstein equations assume the form:
\begin{align}\label{ncstring}
\ddot \Phi + 3H \dot \Phi &=- \frac{\delta V(\Phi)}{\delta \Phi}  + \dots \nonumber \\
3 H^2 &= \left(\frac{\dot \Phi^2}{2} + V(\Phi)\right) + \dots , \nonumber \\
2 \dot H &= - \kappa^2 \, (\rho_\Phi + p_\Phi) = - \dot \Phi^2 + \dots
\end{align}
where the $\dots$ denote non-critical string effects, which may be assumed subdominant in the late epochs, essentially after the exit from inflation, in some non critical string models~\cite{diamand}. In such a case, the dilaton configurations read~\cite{lahanas}
\begin{align}\label{dil}
\Phi(t) = \frac{2}{{\rm c_\Phi}} \, {\rm ln}\Big(\frac{a(t)}{a_I}\Big) + \Phi_I = \frac{2}{{\rm c_\Phi}} {\rm ln}\Big(\frac{{\rm c_\Phi}^2\, H_I}{2} \, (t-t_I) + 1\Big) + \Phi_I,
\end{align}
corresponding to exponential potentials of the form~\cite{diamand,lahanas}
\begin{align}\label{dilpot}
V(\Phi) = \frac{6-{\rm c_\Phi}^2}{2}\, \kappa^{-1}\, H_I\, e^{-{\rm c_\Phi}\, (\Phi - \Phi_I)},
\end{align}
where 
{$c_\Phi$ is some constant}, depending on the details of the non-critical
string model, and the index $I$ denotes values after the exit of inflation. The form of such dilatons implies an evolution for the scale
factor of such a Universe of the form~\cite{lahanas}
\begin{align}\label{eq:scalefactor}
a(t)=a_I \Big( \frac{{\rm c_\Phi}^2\, H_I}{2} \, (t-t_I) + 1\Big)^{\frac{2}{{\rm c_\Phi}^2}}\,,
\end{align}
and a Hubble parameter
\begin{align}\label{hubble}
H^{-1} = H_I^{-1} + \frac{{\rm c_\Phi}^2}{2} \, (t-t_I),
\end{align}
ignoring subdominant  non-critical string effects~\cite{diamand}.  One could then fit such models with the current data to study the alleviation of the tensions, as done in \cite{EPLtensions}, by fitting the phenomenological constant ${\rm c_\Phi}$, appearing in the expression for the potential. As already mentioned, one may even use different $c_\Phi$ at different epochs of the Universe, by allowing for mild time dependence in the parameter $c_\Phi$.}

{However, as mentioned above, such analysis lacks detailed support from microscopic string models, for which the potential is not known, neither the phenomenological constant $c_\Phi$, whose value in general will affect
the phenomenology of strings. Moreover, one needs to reconcile non constant dilatons $\Phi(t)$ with the particle-physics string phenomenology,
given the connection of the string coupling with $e^{\Phi(t)}$, which thus would affect the coupling parameters of the low-energy theory. Although
not impossible, nonetheless such a task is hugely open at present. }

\subsubsection{Quantum-gravity corrections to the Stringy RVM}

{Fortunately, in our stringy RVM there is another route, which is related
to microscopic string models without affecting their phenomenology, in the sense that the dilaton is assumed constant, in agreement with string phenomenology, but the integration of quantum gravitational d.o.f.  (gravitons) is held responsible for logarithmic corrections of the RVM coefficients $\nu_0$ in (approximately) de Sitter phases, such as the space-time characterising the  current era of the Universe~\cite{houston,sugraRVM}.}

{Although we shall leave the detailed analysis for a future publication, we can sketch below the basic idea, which will allow the reader to appreciate better the situation. The analysis resembles that of \cite{houston}, which, although referred to N=1 SUGRA models, with dynamical
supergravity breaking, due to the formation of condensates of gravitino fields as a
consequence of their self interactions, nonetheless it can be easily
adapted to the the stringy RVM case at hand. The idea is that by expanding the metric about a de Sitter space-time, such as the current or inflationary eras, and integrating out graviton degrees of freedom, one arrives at quantum corrected effective actions $\Gamma_{\rm eff}$, in which the de Sitter solution of the space time arises self consistently by looking at minimization of $\Gamma_{\rm eff}$ with respect to the quantum corrected cosmological constant
$\Lambda$: $$\frac{\delta \Gamma_{\rm eff}}{\delta \Lambda }=0~.$$
In late eras of the Universe, due to the smallness of the current-era cosmological constant and the weakness of gravity, one loop corrections are sufficient. As in the N=1 SUGRA case~\cite{houston,fradkin}, reviewed in section \ref{sec:instsugra} above,  the {calculations are performed}  in a Euclidean space-time and at the end one performs analytic continuation  back to Minkowski spacetime.}

{ The one-loop effective action (after performing the graviton integration in a (gauge fixed) {\it Euclidean} quantum-gravity path integral) assumes the generic form \eqref{effpot} where hatted quantities refer to a background space-time, which can be taken to be the (approximately) de-Siiter space time charcterising the current era of the FRLW Universe.
In such cosmological, approximately de Sitter, ({\it Euclidean}) backgrounds,
$$\widehat R_{\lambda\mu\nu\rho} = \frac{\Lambda}{3} \, \Big(\, \widehat g_{\lambda\nu} \, \widehat g_{\mu\rho} - \widehat g_{\lambda\rho} \, \widehat g_{\mu\nu}\Big)~,$$
with $\Lambda = {\rm constant} > 0 $ the quantum-corrections-induced cosmological constant, one has the correspondence \eqref{corr1} ~\cite{houston,sugraRVM}
with $H = H_0$, the approximately constant Hubble parameter at the present era.}

In generic non supersymmetric quantum
gravity RVM models, of potential relevance to the modern era phenomenology of our stringy RVM, where only the graviton d.o.f.  are integrated out, one may use generic tree-level (anti) de Sitter bare
 $\Lambda_0$ as a regulator parameter for the Euclidean quantum-gravity path integrals preserving holography in the sense of an AdS/CFT correspondence~\cite{adscft}, which is known to lead to significant simplifications in obtaining, for instance, stationary black hole solutions in coloured black hole systems, based on SU(N) gauge fields~\cite{negative}.
In such generic  non supersymmetric cases $\Lambda_0$ can be
an arbitrary scale with mass dimension +2. For the validity of perturbation theory and transplanckian conjecture hypothesis, we assume
\begin{align}\label{fkappa}
|\kappa^2 \, \Lambda_0 | \leq 1,
\end{align}
From the form of the coefficients \eqref{coeff}, which also applies to modern eras, as we discussed above,
the ground state of the current-epoch also assumes an RVM like form~\cite{sugraRVM}, augmented, however, with logarithmic
dependences on the one-loop cosmological constant/Hubble parameter  $\Lambda \sim 3H^2$ ($\sim {\rm ln}H^2$) of the RVM coefficients of the vacuum energy density \eqref{rLRVM} .

{From \eqref{effpot}, \eqref{coeff}, we then observe that the quantum corrections induce a renormalization of the gravitational constant~\footnote{{We only
mention that in \cite{houston,sugraRVM} we were interested in examining conditions under which a Starobinsky type inflation would emerge from this
construction. This led to the imposition  of a constraint for the renormalised cosmological constant $\widehat R = 4\Lambda = \Lambda_1$, upon
appropriately fixing the various parameters of the model. In our context of the stringy RVM, inflation is induced by gravitational anomalies, in a
way different from Starobinsky~\cite{staro}, as we have discussed above and in \cite{ms21}, and therefore the above constraint is not imposed. This implies a r\^ole for the one-loop coeffciient  $\alpha_1$ as a correction to the gravitational constant  \eqref{corrkappa}.
It should be stressed that
one-loop quantum-gravity
corrections to the effective gravitational action  of the above form also
exist throughout  the evolution of the stringy RVM Universe,
including its pre-RVM-inflationary phase, see section \ref{sec:instsugra}, as well as the RVM-inflationary epoch.
However, such effects are
subdominant to the effects induced by the CP-violating gravitational anomalous couplings of the KR axion field, which lead to GW condensates \eqref{lambda} that are responsible for the RVM inflationary phase.}}
\begin{align}\label{corrkappa}
\kappa^{-2} \to \kappa^{-2} \, (1 + \alpha_1 )={\kappa^{-2} \, (1 + {\cal O}(\ln H))}
\end{align}
{where we have explicitly remarked the mild logarithmic dependence on $H$ exhibited by  the coefficient $\alpha_1$},  as in the case of type II RVM models \eqref{typeII}, mentioned above~\cite{EPLtensions}. Curiously enough, as follows from \eqref{coeff},
the one-loop corrections to the tree-level cosmological constant  (including those due to fermions \cite{houston,sugraRVM}), do not include ${\rm ln} H$ corrections, which implies a slightly different behaviour from the
type II RVM model, as far as the $c_0$ term is concerned.
Nonetheless, the current stringy RVM, augmented with the aforementioned loop corrections in its coefficients, can be fitted to the data so as to obtain information as to what degree, compared with the other models examined in \cite{EPLtensions}, it can alleviate the $H_0$, $\sigma_8$ tensions. We reserve a detailed phenomenological analysis of such quantum corrected models for a future work.}


\section{Conclusions \label{sec:concl}}

In this work we have discussed a complete cosmological evolution of the stringy RVM of the Universe, completing in a non trivial way our earlier studies. We have discussed potential scenarios for the formation of primordial gravitational waves in very early epochs of the
Universe, immediately after the Big Bang, where the dynamics might be dominated by a first hill-top inflation in some string-inspired scenarios where local supersymmetry is broken dynamically via the formation of gravitino condensates. The gravitino becomes massive in such scenarios  {and decouple  after the SUGRA-induced inflationary phase}, while the graviton remains strictly massless. {This is the reason why we consider only the effective action of the (massless)  bosonic part of the string supermultiplet as the starting point to generate the next inflationary step  (the stringy RVM one) .}

The effective potential of the gravitino condensate has a double-well shape, which however could be statistically biased, insofar as the occupation rates of the two vacua are concerned. This results in the formation of unstable domain walls in the early Universe, after the first inflation, whose non-spherical collapse leads to the formation of gravitational waves. The first inflation is responsible for washing off any potential spatial inhomogeneities in the system, thus making the isotropic and
homogeneous approximation used in the study of the formation of the gravitational-wave condensates a valid one. These condensates are formed when the gravitino condensate has relaxed in the lower-lying vacuum of its potential. In such systems, there is a phase at the exit of the first inflation in which the gravitational-wave condensates have not yet formed, but there are gravitational-wave perturbations in the universe. Such a phase is dominated by a stiff gravitational-axion fluid, with the gravitational
axion field being dual to the antisymmetric tensor spin-one field of the massless gravitational string multiplet and coexists with gravitinos in these early eras. {As noted, gravitinos subsequently decouple (prior to the advent of the RVM phase).}

Due to the gravitational anomalies, which the gravitational axion couples
to, the system of axions coupled to gravitational Chern-Simons anomalous terms is characterised by negative energy density, so one may make an analogy of a kind of phantom-matter era, with an equation of state that of RVM, but with negative energy density, see \eqref{bgCS}.  {Since the relation  $p=-\rho>0$ holds exactly for this system, we have called this equation of state `phantom vacuum'.}
However. the formation of the gravitational-wave condensates, which is assumed to occur simultaneously (in cosmological time scales) with the dominance of the gravitational waves in our string-inspired cosmological model,
implies that the energy density of the total system (axion, gravitational
anomalies and condensates) becomes positive as a result of the role of the condensate. The total equation of state of this fluid is that of the running vacuum \eqref{rvmeos},  hence { $p_{\rm total}=-\rho_{\rm total} <0$ with positive total energy density, and  thus globally this phase is not a phantom-vacuum dominance epoch} but a true RVM-dominance era. The gravitational-wave condensate drives the second RVM type inflation, which is dynamical, without the need for an external inflation field, and is due to the non-linear terms in the RVM energy density. {The dominant term of the latter in the early Universe  is proportional to the fourth power of the Hubble parameter and  is brought about  by the expectation value of
the gravitational-anomalies condensate in our stringy framework.}

In these early phases of the stringy RVM universe only ({massless})  gravitational degrees of freedom are present as external fields. The pre-inflationary phase
contains gravitational anomalies. Chiral matter is generated at the end of the second RVM inflationary period, induced by the
condensate of gravitational waves. The chiral matter is held responsible for a cancellation of the primordial gravitational anomalies, as required
for consistency of radiation and matter quantum field theory. Thus the post inflationary cosmology is more or less standard, at least from the point of view of conservation of energy-momentum tensor (we stress however that there is no fundamental issue with diffeomorphism invariance of the anomalous gravitational theory without matter except the gravitational axionic one. As a theory is perfectly well defined, the apparent non conservation of the gravitational-axion stress tensor merely expressing exchange
of energy between the axion and the gravitational field).

During the second RVM inflation, the ground state of the gravitational axion admits a configuration that spontaneously violates Lorentz symmetry, due to its constant-in-cosmological-time rate of change, in the sense of inducing a preferred frame (the cosmological one).
This configuration, which is exclusively due to the primordial gravitational anomaly condensates, induced by gravitational waves,
remains undiluted at the {exit from} the second RVM inflation, and is responsible, during the radiation era, for leptogenesis in theories containing
right-handed neutrinos, and thus eventually baryogenesis.

In such stringy RVM, the gravitational axions  can acquire non-perturbative potentials at late eras of the universe, and thus masses, playing the r\^ole of dark matter components. In modern eras, the gravitational anomalies resurface, since (chiral) matter is becoming once again subdominant,
compared to vacuum energy contributions. This implies an approximately de
Sitter  phase, of RVM type however, which implies in general observable deviations from the $\Lambda$CDM paradigm, such as alleviations of data tensions. In the current work we have made some {speculations about how the
 late time RVM emerging from our framework might reproduce a type II RVM,
 characterised
by logarithmic dependence in today's Hubble parameter $H_0$  corrections to the effective gravitational constant}. Such type II RVM models are
known to contribute significantly to the alleviation of the $H_0$ and $\sigma_8$ tensions.
Such corrections seem to be generic to quantum graviton effects in quantum-field-theoretic RVM models. We hope to come back to a more detailed phenomenology of the stringy RVM in  a future publication.

\section*{Acknowledgements}

The work  of NEM is supported in part by the UK Science and Technology Facilities  research Council (STFC) under the research grant ST/T000759/1. The work of JS has
been partially supported by projects  PID2019-105614GB-C21 and FPA2016-76005-C2-1-P (MINECO, Spain), 2017-SGR-929 (Generalitat de Catalunya) and CEX2019-000918-M (ICCUB). The authors also acknowledge participation in the COST Association Action CA18108 ``{\it Quantum Gravity Phenomenology in
the Multimessenger Approach (QG-MM)}''.

\end{document}